\documentclass[fleqn,usenatbib,useAMS]{mnras}

\usepackage{graphicx}
\usepackage{bm}
\usepackage{amsmath}	
\usepackage{amssymb}	
\usepackage{multicol,multirow}        
\usepackage{bm}		
\usepackage{pdflscape}	
\usepackage{subcaption}
\usepackage{tabularx}
\usepackage{ulem}

\usepackage[T1]{fontenc}
\usepackage{ae,aecompl}
\usepackage{newtxtext,newtxmath}

\title[Peculiar velocity corrections]{Peculiar velocities in the local Universe: comparison of different models and the implications for $H_0$ and dark matter}

\author[S. S. Boruah, M. J. Hudson and G. Lavaux]{Supranta S. Boruah$^{1, 2, 3}$\thanks{Contact e-mail: \href{mailto:ssarmabo@arizona.edu}{ssarmabo@arizona.edu}}, Michael J. Hudson$^{2,4, 5}$\thanks{Contact e-mail: \href{mailto:mike.hudson@uwaterloo.ca}{mike.hudson@uwaterloo.ca}}, Guilhem Lavaux$^{6}$\thanks{Contact e-mail: \href{mailto:guilhem.lavaux@iap.fr}{guilhem.lavaux@iap.fr}}
\\
$^{1}$Department of Applied Mathematics, University of Waterloo,  200, University Ave W, Waterloo, ON N2L 3G1 \\
$^{2}$Waterloo Centre for Astrophysics, University of Waterloo,  200, University Ave W, Waterloo, ON N2L 3G1\\
$^{3}$ Department of Astronomy and Steward Observatory, University of Arizona, 933 N Cherry Ave, Tucson, AZ 85719, USA \\
$^{4}$Department of Physics and Astronomy, University of Waterloo, Waterloo, ON, N2L 3G1, Canada\\
$^{5}$Perimeter Institute for Theoretical Physics, 31 Caroline St N, Waterloo, ON N2L 2Y5\\
$^{6}$CNRS \& Sorbonne Universit\'e, UMR7095, Institut d'Astrophysique de Paris, F-75014, Paris, France 
}

\date{Last updated XXXX; in original form YYYY}

\pubyear{2020}

\begin{document}
\label{firstpage}
\pagerange{\pageref{firstpage}--\pageref{lastpage}}
\maketitle

\newcommand{\diffop}{\mathrm{d}}
\newcommand{\mmat}[1]{{\mathbf{#1}}}
\newcommand{\mP}{\mathcal{P}}
\newcommand{\mM}{\mathcal{M}}
\newcommand{\mL}{\mathcal{L}}
\newcommand{\mvec}[1]{{\bm{#1}}}
\newcommand{\kmsMpc}{km s$^{-1}$ Mpc$^{-1}$}
\begin{abstract}
When measuring the value of the Hubble parameter, $H_0$, it is necessary to know the recession velocity free of the effects of peculiar velocities. In this work, we study different models of peculiar velocity in the local Universe. In particular, we compare models based on density reconstruction from galaxy redshift surveys and kernel smoothing of peculiar velocity data. The velocity field from the density reconstruction is obtained using the 2M++ galaxy redshift compilation, which is compared to two adaptive kernel-smoothed velocity fields: the first obtained from the 6dF Fundamental Plane sample and the other using a Tully-Fisher catalogue obtained by combining SFI++ and 2MTF. We highlight that smoothed velocity fields should be rescaled to obtain unbiased velocity estimates. Comparing the predictions of these models to the observations from a few test sets of peculiar velocity data, obtained from the Second Amendment Supernovae catalogue and the Tully-Fisher catalogues, we find that 2M++ reconstruction provides a better model of the peculiar velocity in the local Universe than the kernel-smoothed peculiar velocity models. We study the impact of peculiar velocities on the measurement of $H_0$ from gravitational waves and megamasers. In doing so, we introduce a probabilistic framework to marginalize over the peculiar velocity corrections along the line-of-sight. For the megamasers, we find $H_0 = 69^{+2.9}_{-2.8}$ \kmsMpc using the 2M++ velocity field. We also study the peculiar velocity of the the galaxy NGC1052-DF2, concluding that a short $\sim$ 13 Mpc distance is not a likely explanation of the anomalously low dark matter fraction of that galaxy.
\end{abstract}




\section{\label{sec:intro} Introduction}

Peculiar velocities, deviations from the regular Hubble flow, are sourced by the gravitational pull of large-scale structures, thus providing the only way to measure growth of structure on large scale in the low redshift Universe \citep{supercal_growth, sn_flows_paper, adams_and_blake_new}. Apart from their use as a probe of cosmological structure growth, one also has to account for the peculiar velocity of galaxies in other studies of cosmology and galaxy formation. For example, peculiar velocity corrections for nearby standard candles and standard sirens are an important step in trying to measure the expansion rate of the Universe. It was noted in \citet{Hui2006} that correlated errors in the redshifts of supernovae, introduced due to peculiar velocities, are important to account for in cosmological analyses. \citet{NeillHudson2007} applied peculiar velocity corrections while inferring the equation-of-state from supernovae, finding a systematic bias of $\Delta w \sim 0.04$ when not corrected for the peculiar velocities. In \citet{Riess2011}, peculiar velocity corrections were applied for the first time in the measurement of the Hubble constant from supernovae. 

Peculiar velocity corrections are especially important given the increasing discrepancy \citep{H0_tension_review} in the value of the Hubble constant, $H_0$, measured using the Cosmic Microwave Background (CMB) \citep{planck_cosmology} and other low-$z$ measurements \citep{shoes_new, H0licow}. The current tension between the measurements from CMB and the low redshift supernovae has been estimated to be $\sim 4.4\sigma$. Other methods such as standard sirens and megamasers, which measure distances without any calibration to the distance ladder, are crucial in resolution of the $H_0$ tension. $H_0$ has already been measured from the first detection of  gravitational waves from a binary neutron star merger, GW170817 \citep{gw170817} yielding a value of $H_0 = 70.0^{+12.0}_{-8.0}$ \kmsMpc  \citep{ligo_H0}. The Megamaser Cosmology Project \citep[hereafter MCP,][]{MasersH0} has also measured distances to six megamasers giving a measurement, $H_0 = 73.9 \pm 3.0$ \kmsMpc.

The value of $H_0$ measured from local distance indicators depends on the peculiar velocity corrections. The observed redshift, $z_{\text{obs}}$, for standard candles and standard sirens gets a contribution from both the recession velocity due to the Hubble flow, $H_0 r$, and the radial peculiar velocity of the object, $v_r$,

\begin{equation}\label{eqn:z_contribution}
    cz_{\text{obs}} \approx H_0 r + v_r.
\end{equation}
Therefore, in order to measure the value of $H_0$, the radial peculiar velocity needs to be subtracted from the observed redshift. The inferred value of $H_0$ can have significant difference depending on the model of velocity corrections \citep[see e.g.][]{MasersH0}. To measure the peculiar velocity, we need accurate measurements of the  redshifts as well as distances. Accurate distances are also important for measuring, e.g., the masses of galaxies, which in turn has implications for understanding the nature of dark matter. 

To measure the peculiar velocity of galaxies, one has to rely on some distance indicator. Commonly used distance tracers for the measurement of peculiar velocity include empirical galaxy scaling relationships such as the Tully-Fisher \citep[TF, ][]{TF_relations} and the Fundamental Plane \citep[FP, ][]{FP1, FP2} relations as well as Type Ia supernovae. The distance measurements from TF and FP relations have $\sim 15$--$25\%$ uncertainty, while Type Ia SNe gives a distance estimate which is accurate to $5$--$10\%$. However, not all galaxies have a distance estimate obtained by one of these methods. Hence we need some method to map out the peculiar velocity field of the nearby Universe. 

In this work, we compare different peculiar velocity fields by comparing their predictions to independent peculiar velocity catalogues. Broadly, we compare the velocity field predicted using density reconstruction and the velocity field predicted using the adaptive kernel smoothing technique. Predicting peculiar velocities  based on density reconstruction has a long history \citep[see e.g., ][]{Kaiser91, Hudson94b}. In this approach, one uses the galaxy redshift surveys to `reconstruct' the density field, which in turn is used to predict the peculiar velocity field in the local Universe. In contrast, the adaptive kernel smoothing method smooths the peculiar velocity data to map out the velocity field of the Universe. This has been used for cosmography with the 6dF \citep{6df_velocity} and the 2MTF \citep{2mtf_cosmography} peculiar velocity surveys. The reconstructed velocity field used in this work is obtained using the 2M++ galaxy redshift compilation \citep{2Mpp_paper} and the adaptive kernel-smoothed velocity fields are obtained by smoothing the 6dF peculiar velocity catalogue and a combined Tully-Fisher catalogue from SFI++ and 2MTF. We use two different methods, a simple comparison of the mean squared error and a forward likelihood method, to compare the different velocity field predictions to the observations.

We also study the impact of different peculiar velocity models on cosmology and galaxy formation. First, we study its impact on the value of $H_0$ measured from gravitational wave standard sirens and megamaser galaxies. Neither of these techniques rely on intermediate distance calibrators, providing new, direct ways of measuring $H_0$. Finally, we study the peculiar velocity of NGC 1052-DF2, a galaxy that has been found to contain little or no dark matter and where it had been argued that a smaller distance to this galaxy can explain the anomaly.

This paper is structured as follows: in section \ref{sec:pv_data}, we describe the peculiar velocity data, which we use for the adaptive kernel smoothing and to test peculiar velocity models. Section \ref{sec:pv_model} describes the two methods for predicting the peculiar velocity fields. We highlight the importance of scaling the smoothed velocity fields by a constant factor to obtain unbiased velocity estimate of the galaxies in section \ref{sec:scaling}. In section \ref{sec:comparison_metric}, the predictions of the peculiar velocity models are compared to the observed peculiar velocity from the test sets. We discuss the implications of the peculiar velocity fields for measurements of $H_0$ in section \ref{sec:H_0}. Finally, we discuss the case of NGC 1052-DF2 in section \ref{sec:ngc1052} before summarising our results in Section \ref{sec:summary}. In Appendix \ref{sec:KS_simulation}, we tested the predictions of the kernel smoothing method with an N-body simulation to determine the smoothing scale for unbiased velocity estimate. Appendix \ref{sec:posterior_ratio} presents the detailed results of the posterior ratios used for comparing the different peculiar velocity models.
\section{Peculiar Velocity data}\label{sec:pv_data}

We use a few different peculiar velocity catalogues in this work. These catalogues serve two purposes: first, as a `tracer sample' to map the velocity field of the local Universe using an adaptive kernel smoothing technique, and second, as test sets to test the predictions of the peculiar velocity models. In this section, we describe the different catalogues we use in this work, their main features and the corresponding data processing required. 

\subsection{6dF peculiar velocity catalogue}\label{ssec:6df}

\begin{figure*}
    \centering
    \begin{subfigure}[t]{0.5\textwidth}
        \centering
        \includegraphics[width=\linewidth]{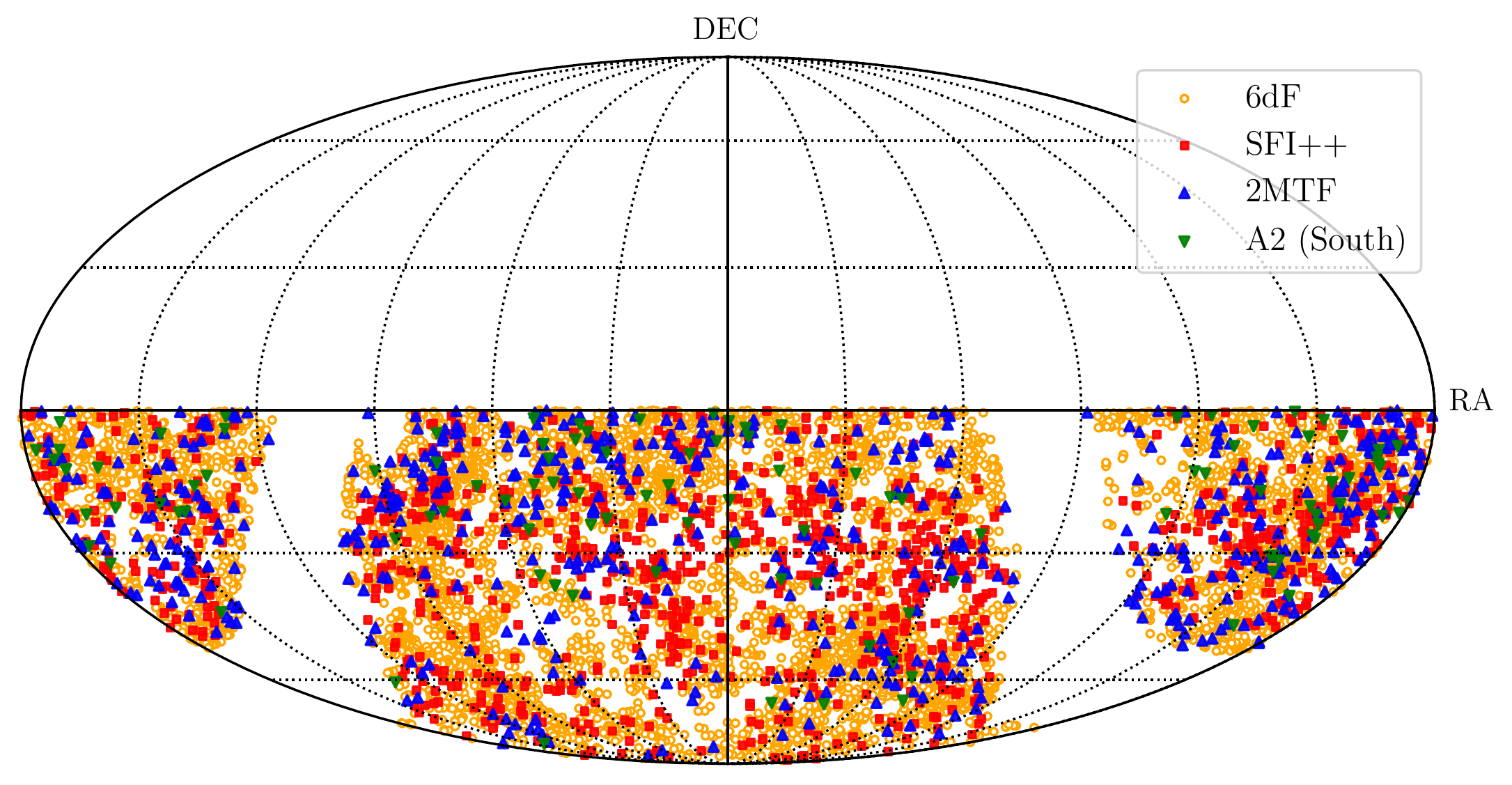}
    \end{subfigure}%
    ~ 
    \begin{subfigure}[t]{0.5\textwidth}
        \centering
        \includegraphics[width=\linewidth]{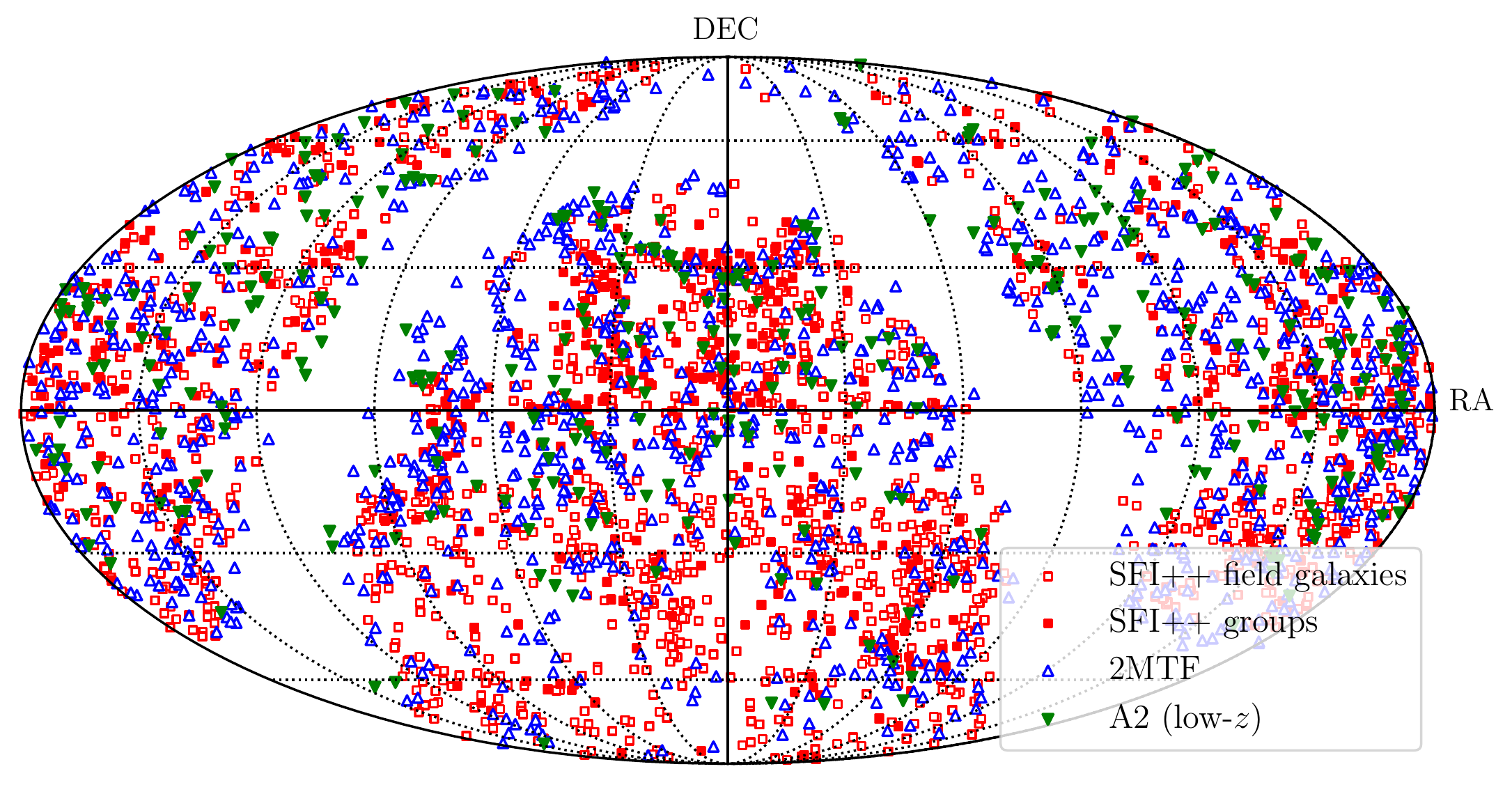}
    \end{subfigure}
    \caption{\textit{Left:} Distribution of 6dF peculiar velocity catalogue objects and the corresponding test objects in the southern hemisphere plotted in the equatorial coordinates. The 6dF objects are shown with orange circle. The test set consists of objects from the SFI++, 2MTF and the A2-South catalogue - shown with a red square, blue triangle and a green inverted triangle respectively. \textit{Right:} The sky distribution of SuperTF and A2 objects plotted in the equatorial coordinates. The SFI++ objects are denoted with a red square with the groups being a filled squared and the field galaxies are hollow squares. The 2MTF galaxies are denoted with a blue triangle and the A2 supernovae as a green inverted triangle.}
    \label{fig:sky_distribution}
\end{figure*}

The 6dF peculiar velocity sample \citep{6df_velocity} consists of galaxies from the fundamental plane survey \citep{6df_fp, 6df_fp2} of the 6dF galaxy survey. It is presently the largest peculiar velocity survey with a total of $8~885$ galaxies. The distance (and hence the radial peculiar velocity) of these galaxies is estimated using the Fundamental Plane relation. The sample is restricted to the southern hemisphere with a galactic cut of $|b| > 10^{\circ}$ and $cz < 16~000$ km/s in the CMB frame. The mean distance uncertainty of the sample was found to be $\sim 26 \%$. We plot the sky distribution of the 6dF peculiar velocity catalogue along with the tracers we use to compare the velocity field in the southern hemisphere in the left panel of Figure \ref{fig:sky_distribution}. We use the 6dF peculiar velocity sample as a tracer sample to predict the peculiar velocity using the adaptive kernel smoothing method described in Section \ref{ssec:aks}.

\subsection{Tully-Fisher catalogues}

In this work, we use two Tully-Fisher (TF) catalogues: SFI++ and 2MTF. The TF catalogues serve dual purposes in this work. First, we use the objects in the southern hemisphere as a test set to test the predictions of the 2M++ reconstructed velocity field and the adaptive kernel smoothed velocity obtained from 6dF. Second, we use a combined TF catalogue of SFI++ and 2MTF to predict the velocity field using the adaptive kernel smoothing method and compare it with the other velocity field models.

\subsubsection{SFI++}
SFI++ \citep{sfi1, sfi2} is an $I$-band Tully-Fisher survey with more than $4000$ peculiar velocity measurements. As noted in \citet{sn_flows_paper}, there is a deviation from the linear Tully-Fisher relationship in both the faint and bright end. We therefore only use galaxies with $-0.1 < \eta < 0.2$, where, $\eta$ is related to the velocity width, $W$, as $\eta = \log_{10}W - 2.5$. To remove the outliers, we iteratively fit the Tully Fisher relation using the redshift space distances and remove the $3.5\sigma$ outliers. For this work, while comparing the peculiar velocities, we only consider the galaxies that are within $cz < 10000$ km/s. With these cuts, we have a total of $1607$ field galaxies and $584$ groups in our sample. Of these, $949$ field galaxies and $204$ groups are in the southern hemisphere. While comparing the 6dF adaptive kernel-smoothed peculiar velocity field and the 2M++ reconstructed velocity field, we use only the galaxies in the southern hemisphere as a test set.

\subsubsection{2MTF}
The 2MTF survey \citep{2MTF} is a Tully-Fisher survey in the $J, H$ and $K$ bands. The final catalogue \citep{2mtf_data} consists of 2062 galaxies within $cz < 10000$ km/s. We remove the duplicates from 2MTF that are already contained in the SFI++ catalogue. Similar to the SFI++ catalogue, we only use galaxies with $-0.1 < \eta < 0.2$ and reject outliers by iteratively fitting the Tully-Fisher relation. We have a total of $1248$ galaxies after these cuts. Of these, $567$ galaxies are in the southern hemisphere and we use these galaxies to compare with the 6dF adaptive smoothed peculiar velocity field and the 2M++ reconstructed velocity field.

\subsubsection{SuperTF}

\begin{figure}
    \centering
    \includegraphics[width=\linewidth]{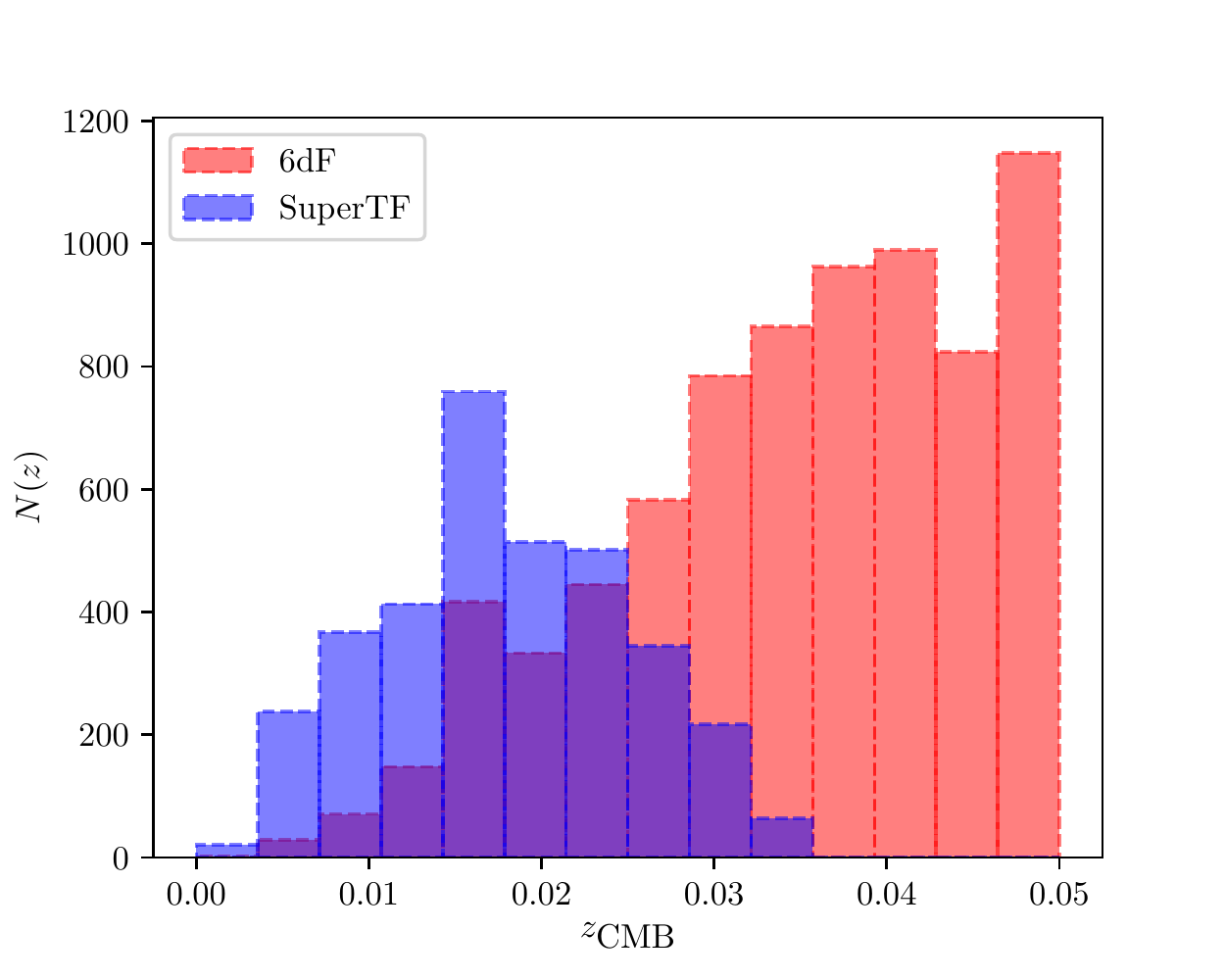}
    \caption{Redshift distribution of SuperTF and 6dF objects. Note that the two catalogues cover different fractions of the sky, with 6dF covering only the southern hemisphere. Also note that the SuperTF catalogue has a higher density of objects compared to 6dF at lower redshifts $(z \lesssim 0.015)$.}
    \label{fig:z_dist}
\end{figure}

\begin{figure*}
    \centering
    \includegraphics[width=\linewidth]{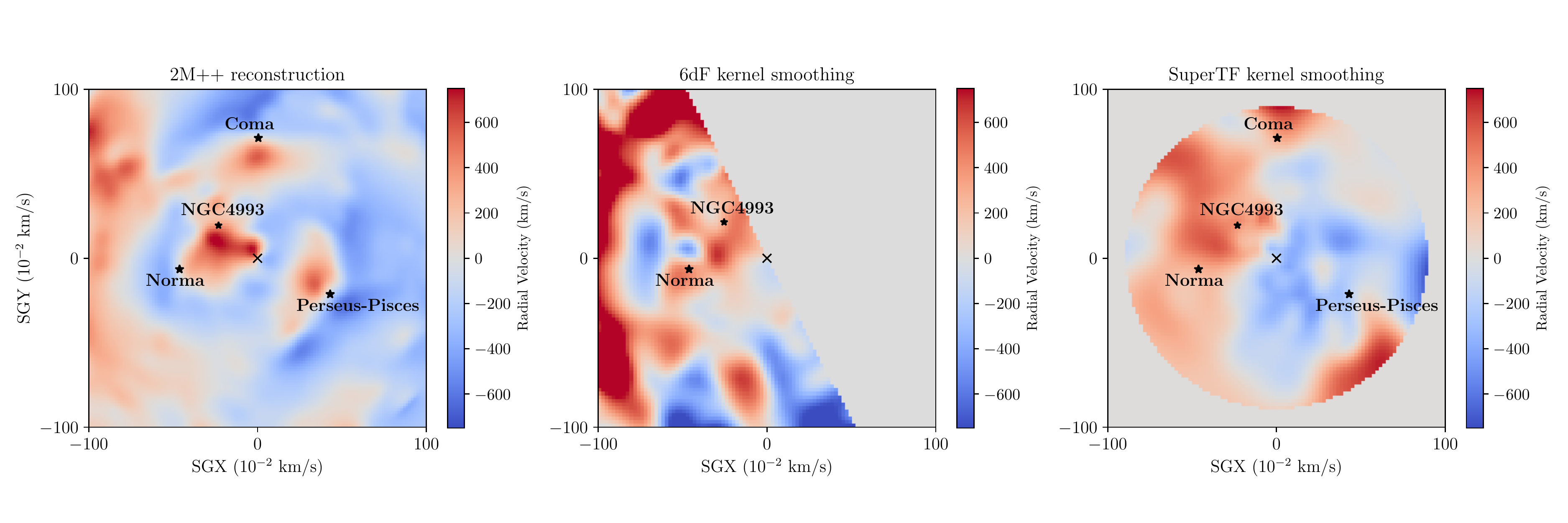}
    \caption{The radial velocity in the supergalactic plane for the 3 different peculiar velocity models: {\it Left:} 2M++ reconstruction, {\it Centre:} Adaptive kernel smoothing with the 6dF peculiar velocity catalogue, {\it Right:} Adaptive kernel smoothing with the SuperTF catalogue. The coordinates for the adaptive smoothed fields are in redshift space, while that for 2M++ reconstruction is in the real space. We also show the location of a few prominent superclusters and NGC4993. }
    \label{fig:sg_velocity}
\end{figure*}

 We combine the SFI++ and the 2MTF catalogs into a `super TF' catalog which we then use to produce an adaptive kernel-smoothed peculiar velocity map. Unlike 6dF, we can use this catalogue to map out the velocity field in both hemispheres using the kernel smoothing method. The $I$-band Tully-Fisher relation, used in the SFI++ catalogue, has a smaller intrinsic scatter compared to the TF relation in the infrared bands, which is used by the 2MTF survey \citep[see e.g., ][]{sn_flows_paper}. Therefore, when there are duplicates in the SFI++ and 2MTF datasets, we use the SFI++ objects. The final data set consists of $584$ SFI++ groups, $1607$ SFI++ field galaxies and $1248$ 2MTF galaxies. We show the sky distribution of the objects in this combined catalogue in Figure \ref{fig:sky_distribution}. We also compare the redshift distribution of the objects in the SuperTF catalogue with the 6dF peculiar velocity catalogue in Figure \ref{fig:z_dist}. Note that the SuperTF catalogue has a higher density of objects at lower redshifts ($z \lesssim 0.015$). Also note that for predicting the velocity using adaptive kernel smoothing, we do not impose any redshift cuts on the data sets. The velocity field of the local Universe mapped using the adaptive kernel smoothing technique on the SuperTF catalogue is shown in Figure \ref{fig:sg_velocity}.

\subsection{A2 Supernovae}

We also use a sample of Type Ia supernovae to test the peculiar velocity fields. Type Ia Supernovae are excellent distance indicators with much smaller distance error than TF or FP galaxies. We presented the Second Amendment (A2) sample of supernovae in \citet{sn_flows_paper}. It consists of low redshift (low-$z$) supernovae from the CfA supernovae sample \citep{constitution}, Carnegie Supernovae project \citep{CSP_sn1_folatelli, CSP_DR3}, the Lick Observatory Supernova Survey (LOSS) \citep{LOSS_data} and the Foundation supernovae sample \citep{foundation1, foundation2}. The final sample consists of 465 low-$z$ supernovae, resulting in the largest peculiar velocity catalogue based on supernovae. While comparing to the 6dF adaptive smoothed peculiar velocity field, we only use the SNe in the southern hemisphere and $|b| < 10 ^{\circ}$. We call this data set consisting of 150 SNe, `A2-South'.

\section{Peculiar velocity models}\label{sec:pv_model}

In this work, we compare two different types of peculiar velocity models - {\it i)} the velocity field reconstructed from the 2M++ galaxy redshift compilation, {\it ii)} velocity field mapped out using an adaptive kernel smoothing technique. In this section, we briefly describe both these methods.

\subsection{Velocity reconstruction with 2M++}\label{ssec:reconstruction_scheme}

We use the velocity field reconstructed using the 2M++ compilation of galaxy redshifts \citep{2Mpp_paper}. The 2M++ catalogue consists of a total of $69160$ galaxies. The catalogue was found to be highly complete up to a distance of $200\ h^{-1}$ Mpc (or $K < 12.5$) for the region covered by the 6dF and SDSS and up to $125\ h^{-1}$ Mpc (or $K < 11.5$) for the region that is not covered by these surveys. 

We use the luminosity-weighted density field from \citet{Carrick_et_al} in this work. The velocity field is predicted from the density field using linear perturbation theory. The predicted velocity field from the luminosity-weighted density field is scaled by a factor of $\beta = f/b$ and an external velocity, $\mvec{V}_{\text{ext}}$ is added to this. To fit for the value of $\beta$ and the external velocity $\mvec{V}_{\text{ext}}$, we compared the predicted velocity to the observed velocities from the SFI++ and the A2 catalogue. The details of the fitting process can be found in \citet{Carrick_et_al} and \citet{sn_flows_paper}. More details on the reconstruction procedure can be found in \citet{Carrick_et_al}.

\subsection{Adaptive kernel smoothing method}\label{ssec:aks}

An adaptive kernel smoothing technique to map the peculiar velocity field using measured peculiar velocities was presented in \citet{6df_velocity, 2mtf_cosmography}. We use this method to map the velocity field in the local Universe using the 6dF and the SuperTF catalogues.

In this scheme, a Gaussian kernel is used to smooth the peculiar velocity measurements from the catalogues. The measured radial velocities, $\{v_r(\mvec{r}_i)\}$, in peculiar velocity catalogue at locations, $\{\mvec{r}_i\}$, is used to predict the peculiar velocity at $\mvec{r}$ as\footnote{Note that there are other ways of interpolating between observed radial velocities. For example, \citet{Dekel1990} suggested the use of a tensor window function to smooth the observed velocities in order to obtain unbiased velocity estimate. Such window functions may perform better than the one used in this work. However, the effect of these extended window functions are beyond the scope of this work and needs further exploration.}, 

\begin{equation}\label{eqn:kernel_smoothing_v}
    v(\mvec{r}) = \frac{\sum_{i=1}^{N_{\text{gal}}} v_r(\mvec{r}_i) \cos \theta_i e^{-\Delta r^2_i/2\sigma_i^2} \sigma_i^{-3}}{\sum_{i=1}^{N_{\text{gal}}} e^{-\Delta r^2_i/2\sigma_i^2}\sigma_i^{-3}},
\end{equation}
where, $\Delta r_i = |\mvec{r} - \mvec{r}_i|$ and $\cos(\theta_i) = \hat{\mvec{r}}\cdot\hat{\mvec{r}}_i$. Note that in this method, the location, $\mvec{r}$, is given in the redshift space. The kernel width for each galaxy is adaptively computed using the prescription presented in \citet{6df_velocity, 2mtf_cosmography}. Assuming a fiducial smoothing length, $\sigma^{\prime}$, the adaptive smoothing length for each object is defined as, 

\begin{equation}\label{eqn:adaptive_sigma}
    \sigma_i = \sigma^{\prime}\left[\frac{\exp\left(\sum_{j=1}^N\ln(\delta_j)/N\right)}{\delta_i}\right]^{1/2},
\end{equation}
where $\delta_i$ is computed as, 

\begin{equation}\label{eqn:delta_def}
    \delta_i = \sum_{j=1}^{N_j} \exp\bigg(-\frac{|\mvec{r}_j - \mvec{r}_i|^2}{2 \sigma^{\prime 2}}\bigg).
\end{equation}
The sum in equation \eqref{eqn:delta_def} is over the $N_j$ objects that are within distance,  $3\sigma^{\prime}$ of the $i$-th object. The quantity, $\delta_i$ roughly calculates the density of peculiar velocity tracers near the $i$-th object. Equation \eqref{eqn:adaptive_sigma} then calculates the kernel width adaptively based on the density of the peculiar velocity tracers. The calculated kernel size in regions with larger density is thus smaller.

\begin{figure}
	\includegraphics[width=\linewidth]{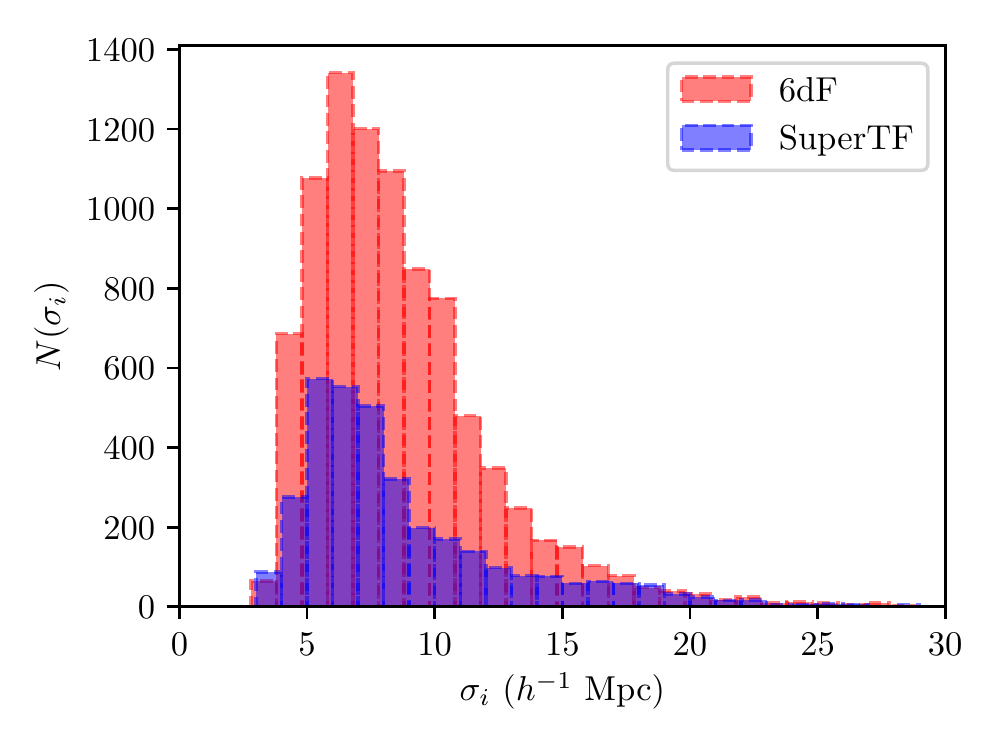}
	\caption{Distribution of the adaptively calculated kernel smoothing length computed using a fiducial smoothing length, $\sigma^{\prime} = 8~h^{-1}$ Mpc, for the two peculiar velocity catalogues, 6dF and SuperTF. The mean smoothing lengths for the 6dF and the SuperTF catalogues are $8.61~h^{-1}$ Mpc and $8.83~h^{-1}$ Mpc respectively.}
	\label{fig:adaptive_sigma}
\end{figure}

We use the adaptive kernel smoothing method to predict the velocity field using the 6dF Fundamental Plane and the SuperTF catalogue. In this work, we use two different fiducial smoothing lengths, $8 h^{-1}$ Mpc and $16 h^{-1}$ Mpc to predict the kernel smoothed velocity field. If we use a fiducial smoothing length of $\sigma^{\prime} = 8~h^{-1}$ Mpc, the mean smoothing length, $\langle \sigma_i \rangle$, for the 6dF and the SuperTF catalogues are $8.61~h^{-1}$ Mpc and $8.83~h^{-1}$ Mpc respectively. The spread of the same smoothing lengths as measured using the standard deviation of the distribution are, $3.60~h^{-1}$ Mpc and $4.39~h^{-1}$ Mpc for the 6dF and Super TF catalogues respectively. The distribution of the adaptively calculated smoothing lengths with fiducial smoothing length of $8 h^{-1}$ Mpc for the 6dF and the SuperTF catalogues is shown in Figure \ref{fig:adaptive_sigma}.

A comparison of the peculiar velocity fields in the supergalactic plane derived using the 2M++ reconstruction, and the adaptive kernel smoothing on the 6dF and the SuperTF catalog is shown in in Figure \ref{fig:sg_velocity}. 

\section{Scaling the smoothed velocity}\label{sec:scaling}

In the previous section, we presented two ways to predict the velocities of galaxies. Both these methods rely on smoothing the peculiar velocity field in some way. However, one must be careful while using a smoothed field to predict velocities since certain smoothing scales may lead to biased estimates of the peculiar velocity. In \citet{unbiased_smoothing} and \citet{Carrick_et_al}, the effect of the smoothing radius on the inferred value of $\beta$ (or $\Omega_m$ in the earlier paper) was studied in the context of constraining cosmological parameters from density-velocity comparison. Using tests on numerical simulations, it was found in \citet{Carrick_et_al} that a smoothing length of $\sim 4 h^{-1}$ Mpc gives unbiased estimates for the peculiar velocity. Using a smoothing length other than this value may bias our estimates of peculiar velocity. Therefore, when predicting the peculiar velocity using the 2M++ reconstruction, we smoothed the velocity field using a Gaussian filter with a smoothing length of $4 h^{-1}$ Mpc to predict the peculiar velocity.


The kernel smoothing method also predicts the velocity by smoothing the peculiar velocity data in the neighbouring region. Given the large error bars and the sparse sample of peculiar velocity data, smoothing over a larger region may be necessary to reduce the uncertainties on the predictions to an acceptable level. However, smoothing over larger regions also biases low the peculiar velocity estimates. Therefore, if we smooth the peculiar velocity field using a smoothing scale $\gtrsim 4 h^{-1}$ Mpc, we need to correct for this bias. To do so, we use a simple scaling of the smoothed peculiar velocity. More precisely, the peculiar velocity estimate, $v_p(R)$ obtained by using a smoothing scale, $R$ is scaled by factor of $A(R)$ such that
\begin{equation}
    v_p(R) \rightarrow A(R) v_p(R).
\end{equation}
The scaling factor, $A(R)$ is determined by comparing the kernel-smoothed velocity of an N-body simulation to the true velocity of the halos. The details of the study on the simulation is presented in Appendix \ref{sec:KS_simulation}. With the help of these simulations, we determine the scaling factor for a smoothing length of $8~h^{-1}$ Mpc and $16~h^{-1}$ Mpc to be $1.07$ and $1.16$ respectively. We compare both the scaled and the unscaled versions of the adaptive kernel-smoothed peculiar velocity fields in the next section.


\section{Comparing peculiar velocity models}\label{sec:comparison_metric}

To compare the models of peculiar velocity of the local Universe, we use independent peculiar velocity data sets to compare the predictions of the models to observations. We propose two ways to test this - the first method is based on a simple comparison of the mean squared error between predicted and observed peculiar velocity of the tracer peculiar velocity data set. In this method, we use the velocity estimate from the peculiar velocity models at the estimated distance of the peculiar velocity tracer and compare it with the observed value of the velocity. However, this method is known to be affected by inhomogeneous Malmquist bias. The second method is the {\it Forward likelihood} method which can correct for the inhomogeneous Malmquist bias. We use Bayesian model comparison with this method to compare the peculiar velocity models presented in Section \ref{sec:pv_model}.

\subsection{Comparing the mean squared error}\label{ssec:mse}

\begin{figure*}
    \centering
    \includegraphics[width=\linewidth]{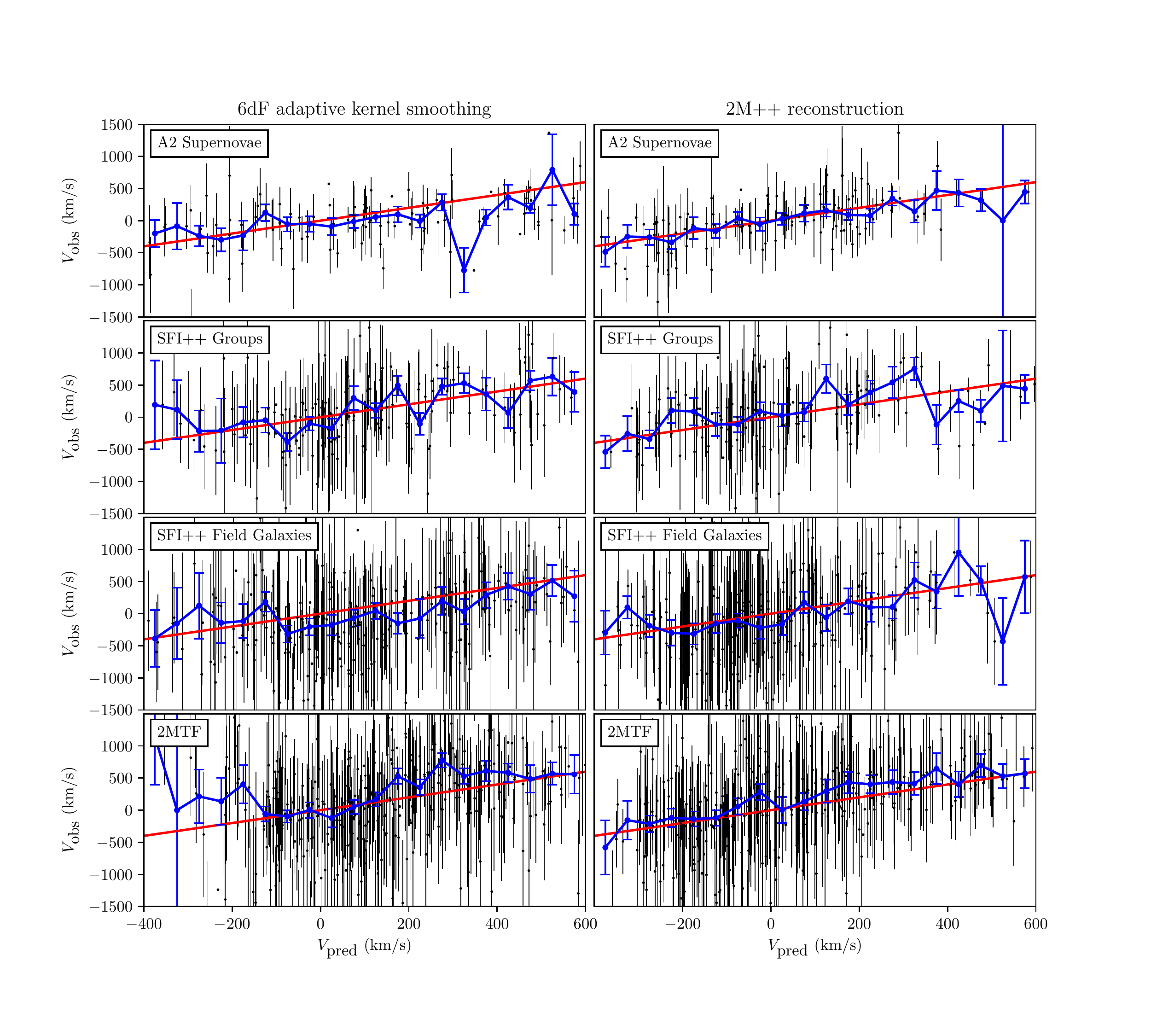}
    \caption{Plot of the predicted velocity, $V_{\text{pred}}$, predicted using adaptive kernel smoothing on 6dF ({\it left}) and 2M++ reconstruction ({\it right}) vs the observed velocity, $V_{\text{obs}}$ for the different test sets. The black markers denote the peculiar velocity estimate for each object in the test set. The blue curve shows a binned version where we plot the weighted average of the observed peculiar velocity objects in each $50$ km s$^{-1}$ bins of predicted peculiar velocity. The red line represents, $V_{\text{pred}} = V_{\text{obs}}$.}
    \label{fig:v_pred}
\end{figure*}

In the first approach, we calculate the mean squared error between the predicted radial velocity from the peculiar velocity models and the measured radial peculiar velocity of the test set. The model with the better predictions for the peculiar velocity should have a lower value of mean squared error (MSE). For the adaptive kernel smoothing approach, we estimate the peculiar velocity of the test objects using the approach of Section \ref{ssec:aks}. In this approach, we use the redshift space position of the test object to estimate the velocity. For the reconstructed velocity field, we predict the velocity at the reported mean position of the object. We plot the predicted velocity from 6dF adaptive kernel smoothing technique and from the 2M++ reconstruction against the observed velocity of the 2MTF, SFI++ and the A2 supernovae in Figure \ref{fig:v_pred}. 

We then compare the estimated velocity with the measured radial velocity. The mean squared error (MSE) between the observed velocity and the predicted velocity is defined as, 

\begin{equation}
    \text{MSE} = \frac{1}{N_{\text{tracers}}}\sum_{i=0}^{N_{\text{tracers}}}\frac{(V^r_{\text{pred}} - V^r_{\text{obs}})^2}{\Delta V^{2}_{\text{obs}}}
\end{equation}
The value of MSE obtained for the different models is presented in Table \ref{tbl:chi_squared}. Since the SuperTF catalogue consists of SFI++ and 2MTF galaxies, we do not use these data sets to compare to the velocity field obtained by using the adaptive kernel smoothing on the SuperTF catalogue. We notice that for all the test sets, the 2M++ reconstructed velocity field gives a lower value of the MSE than the adaptive kernel-smoothed velocity fields. Hence, it suggests that 2M++ reconstructed peculiar velocity is a better model for the peculiar velocity of the local Universe than the one obtained by using adaptive kernel smoothing.

\begin{table}
  \centering
  \caption{The value of MSE measured for different test data sets}
  \begin{tabular}{l c c c}
  \hline
  \multirow{2}{*}{Test} & \multicolumn{3}{c}{Velocity field model} \\
      & 2M++ & 6dF & SuperTF \\
    \hline
   A2 (South) & $0.742$ & $1.856$ & $1.970$\\
   A2 (low-$z$) & $0.899$ & --- & $2.214$\\
    SFI++ groups & $0.814$ & $0.949$ & --- \\
    SFI++ field galaxies & $0.703$ & $0.763$ & --- \\
    2MTF & $0.875$ & $1.072$ & --- \\
    \hline
  \end{tabular}
  \label{tbl:chi_squared}
\end{table}

\subsection{Forward likelihood}\label{ssec:fwd_lkl}

The approach used in Section \ref{ssec:mse} is simplistic and is known to be affected by inhomogeneous Malmquist bias. \citet{Hudson94a, Hudson94b} introduced an approach for peculiar velocity analysis that can deal with inhomogeneous Malmquist bias by using an improved radial distribution for the peculiar velocity tracer \citep[see also][]{pike_hudson}. This approach is called \textit{Forward likelihood}. In this section, we use this approach to determine which peculiar velocity field fits the data well. 

In the forward likelihood approach, the difference in the observed and predicted redshifts are minimized along the line of sight and the radial distribution is marginalized. We take into account the inhomogeneities along the line of sight to correct for inhomogeneous Malmquist bias. This is done by assuming the following radial distribution.
\begin{equation}\label{eqn:r_prior}
	\mathcal{P}(r) = \frac{r^2 \exp\bigg(-\frac{(r - d)^2}{2\sigma^2_d}\bigg) (1 + \delta_g(\mvec{r}))}{\int_0^{\infty} \diffop r^{\prime} r^{\prime 2} \exp\bigg(-\frac{(r^{\prime} - d)^2}{2\sigma^2_d}\bigg) (1 + \delta_g(\mvec{r^{\prime}}))}\;,
\end{equation}
where, $d$ is the distance reported in the peculiar velocity survey converted to comoving distance, $\sigma_d$ is the associated uncertainty, and $\delta_g$ is the overdensity in the galaxy field. In this work, we use the luminosity-weighted density field from the 2M++ reconstruction for the inhomogeneous Malmquist bias correction.

To account for the errors that arise because of the triple-valued regions and inhomogeneities along the line of sight, the likelihood is marginalized over the above radial distribution. The likelihood for observing the redshift given a peculiar velocity model, $\mvec{v}$, can be written as

\begin{equation}\label{eqn:fwd_lkl}
	\mathcal{P}(z_{\text{obs}}|\mvec{v}) = \int_0^{\infty} dr \mathcal{P}(z_{\text{obs}}| r, \mvec{v}) \mathcal{P}(r),
\end{equation}
where, $\mathcal{P}(z_{\text{obs}}|r, \mvec{v})$ is modelled as a Gaussian with standard deviation, $\sigma_v$,
\begin{equation}\label{eqn:conditional_prob_z}
	\mathcal{P}(z_{\text{obs}}|r, \mvec{v})  = \frac{1}{\sqrt{2\pi \sigma^2_v}}\exp\bigg(-\frac{(cz_{\text{obs}} - c z_{\text{pred}}(r, \mvec{v}))^2}{2\sigma^2_v}\bigg)\,,
\end{equation}
and $\mathcal{P}(r)$ is given by Equation~\eqref{eqn:r_prior}. $z_{\text{pred}} \equiv z_{\text{pred}}(r, \mvec{v})$ as given as,

\begin{equation}\label{eqn:z_pred}
    1 + z_{\text{pred}} = \bigg(1 + z_{\text{cos}}(r)\bigg)\bigg(1 + \frac{v_r}{c}\bigg)\,,
\end{equation}
where, $v_r$ is predicted using the velocity model, $\mvec{v}$, at the position, $\mvec{r}$. In the case of adaptive kernel smoothing, the radial velocity is calculated in the redshift space. Therefore, we introduce the redshift space coordinate, $\mvec{s}$, to facilitate the calculation for this case. 

\begin{equation}\label{eqn:redshift_coord}
    \mvec{s} = H_0 \mvec{r} + v_r(\mvec{r})\hat{\mvec{r}}.
\end{equation}
Note that, in the above equation distance coordinates, $\mvec{s}$ and $\mvec{r}$ has the same units as that for the velocity. That is, we convert the distance units into velocity units (km s$^{-1}$). 
We use this change of coordinates to calculate Equation \eqref{eqn:fwd_lkl} in the redshift space,

\begin{align}\label{eqn:fwd_lkl_s}
	\mathcal{P}(z_{\text{obs}}|\mvec{v}) &= \int_0^{\infty} ds \bigg|\frac{\partial r}{\partial s}\bigg| \mathcal{P}(z_{\text{obs}}| r(\mvec{s}), \mvec{v}(\mvec{s})) \mathcal{P}(r(s)), \nonumber \\
	&= \int_0^{\infty} \frac{ds}{H_0} \bigg( 1 - \frac{\partial v_r}{\partial s}\bigg) \mathcal{P}(z_{\text{obs}}| r(\mvec{s}), \mvec{v}(\mvec{s})) \mathcal{P}(r(s))
\end{align}
Note that the above relation is defined only if the relation between $\mvec{r}$ and $\mvec{s}$ is monotonic along the line of sight of the peculiar velocity tracers. As is well-known, this is not always the case due to the phenomenon of triple-valued regions \citep{pec_vel_review}. A triple valued region usually occurs in the neighborhood of big clusters, where, for a given value of redshift, there are 3 solutions in the real space. Since the adaptive    kernel smoothing technique estimates the velocity in the redshift space, it does not take into account the effect of triple-valued points. In this work, to convert between $\mvec{r}$ and $\mvec{s}$ we use the peculiar velocity models to get the peculiar velocity at a given location. We only consider for comparison the objects which have a monotonic relation between $\mvec{r}$ and $\mvec{s}$ under all the peculiar velocity models under consideration. For such objects, Equation \eqref{eqn:fwd_lkl_s} gives a valid way to calculate the forward likelihood using the adaptive kernel smoothed peculiar velocity field. 

\subsection{Bayesian model comparison with Forward Likelihood}

We use Bayesian model comparison to compare the two peculiar velocity models described in Section
\ref{sec:pv_model}. Given two models, $\mM_1, \mM_2$ describing the same data, $D$, Bayesian model comparison gives a way to compare the two models. The plausibility of the two models can be compared by calculating the posterior probability ratio \citep{mackay_book}, 
\begin{equation}
    \frac{\mP(\mM_1|D)}{\mP(\mM_2|D)} = \frac{\mP(\mM_1)}{\mP(\mM_2)} \frac{\mP(D|\mM_1)}{\mP(D|\mM_2)}.
\end{equation}
$\mP(\mM)$ denotes the prior belief in the model, $\mM$. 

Hence, Bayesian model comparison is well-suited to compare the different peculiar velocity models. We do so by using the forward likelihood to calculate the likelihood, $\mP(D|\mvec{v})$. If we assign equal prior probability to two different peculiar velocity models, $\mvec{v}_1$ and $\mvec{v}_2$, i.e., $\mP(\mvec{v}_1) = \mP(\mvec{v}_2)$, the posterior probability ratio can be written as, 

\begin{equation}\label{eqn:post_ratio_v}
    \frac{\mP(\mvec{v}_1|D)}{\mP(\mvec{v}_2|D)} = \frac{\mP(D|\mvec{v}_1)}{\mP(D|\mvec{v}_2)}.
\end{equation}
The right hand side of equation \eqref{eqn:post_ratio_v} can be calculated using equations \eqref{eqn:fwd_lkl} and \eqref{eqn:fwd_lkl_s}. We fix the value of $\sigma_v$ to $150$ km s$^{-1}$.  We present the values for the ratio $\mP(D|\mvec{v})$ for the 2M++ reconstructed velocity field and the kernel smoothed velocity field in Table \ref{tbl:fwd_lkl} and \ref{tbl:fwd_lkl_16}. The fiducial smoothing radius was chosen as $8 h^{-1}$ Mpc and $16 h^{-1}$ Mpc for Table \ref{tbl:fwd_lkl} and \ref{tbl:fwd_lkl_16} respectively. We highlight here some of the main findings:

\begin{itemize}
    \item[{\it i)}] We find that the 2M++ reconstructed velocity field gives a better fit to the observed velocities for all test sets and for all range of redshifts than the adaptive kernel-smoothed velocity fields.
    \item[{\it ii)}] We also find that choosing a bigger smoothing scale of $16 h^{-1}$ Mpc for the adaptive kernel-smoothed velocity performs better than a smoothing scale of $8 h^{-1}$ Mpc for almost all test data sets and all range of redshifts.
    \item[{\it iii)}] The effect of scaling the predicted velocity field as we proposed in Section \ref{sec:scaling} is more pronounced when we use kernel smoothing with a smoothing radius of $16~h^{-1}$ Mpc. The value of $A$ for the two smoothing scales are $1.12$ and $1.34$ for $8h^{-1}$ Mpc and $16 h^{-1}$ Mpc respectively. For the 2MTF, SFI++ field galaxies and SFI++ groups data sets, the scaled velocity from kernel smoothing indeed gives a better fit compared to the unscaled version. However, for the supernovae data, the reverse is true. 
\end{itemize}

\subsection{Comparison of the reconstruction based and kernel smoothing method}\label{ssec:compare_aks_rec}

We saw in this section that the 2M++ reconstructed velocity field gives a better fit to the different data sets compared to the velocity field calculated using adaptive kernel smoothing. In this section, we give a rough argument for why that is the case. 

In Equation \eqref{eqn:kernel_smoothing_v}, if $N$ objects contribute significantly to the calculation of the kernel-smoothed velocity, and if the velocity error for each object is a constant, $\Delta V$, the error in the estimated velocity will be roughly, $\sim \Delta V/\sqrt{N}$. The relative error on the peculiar velocity measured using TF or FP is usually $15$-$25\%$ of the distance given in km/s. For NGC4993, which is at $\sim 3000$ km/s, assuming a relative error of $20\%$, and $N \sim 10$, which is the number of 6dF objects within $8~h^{-1}$ Mpc (number of SuperTF objects within the same distance is $11$), we get the velocity error of $\sim~180$ km/s. This estimate is close to the estimated error, $153$ km/s calculated for the 6dF velocity field. However, the same argument also implies that the error on the calculated peculiar velocity field grows with distance. For redshifts much larger than that of NGC4993, one would need a much higher density of peculiar velocity tracers to achieve a distance error of $\sim 150$ km/s using the kernel smoothing technique. On the other hand, the error with the predictions of the reconstruction based approach arises from systematic error in the reconstruction and errors in the prediction from linear perturbation theory. While some dependence of this error with distance is expected, it is not as drastic as using the kernel smoothing method.

Another problem with the kernel smoothing technique is that it does not account for triple valued regions. Triple valued regions arises when the same value of redshift has 3 solutions in the real space. To get these solution, these points in the real space must have peculiar velocities in opposing directions. Since kernel smoothing is done in the redshift space, potentially these triple valued regions are smoothed together, hence, inducing additional biases.
\section{Implications for \texorpdfstring{$H_0$}{}}\label{sec:H_0}

As we saw in Equation \eqref{eqn:z_contribution}, the redshift of a galaxy gets a contribution from both the peculiar velocity and the Hubble recession. Therefore, one needs to correct for the contribution of the peculiar velocity in order to correctly infer $H_0$. In this section, we consider the peculiar velocity corrections for two data sets: the distance measurement from gravitational wave for NGC4993 and the peculiar velocity corrections for megamasers from the Megamaser Cosmology Project (MCP). We begin with the Bayesian model we use for the treatment of peculiar velocity for these data sets. 

\subsection{\texorpdfstring{$H_0$}{} likelihood}\label{ssec:H0_lkl}

When correcting the redshifts for the peculiar velocities, traditionally one uses a point estimate for the peculiar velocity at a given redshift. However, given the large uncertainty on the distances, we need to marginalize over the radial velocity along the line-of-sight for each object. In this section, we derive the likelihood for such a method. 

We want to infer the Hubble constant, $H_0$, given the observed redshift to the objects, $\{z_{\text{obs}}\}$, and some data, $\{x_{\text{dist}}\}$, from which the distance is derived. That is, we want to calculate $\mP(H_0|\{z_{\text{obs}}\}, \{x_{\text{dist}}\})$. Using Bayes theorem and assuming the different distance measurements are independent, we can simplify the posterior as,

\begin{align}
    \mP(H_0|\{z_{\text{obs}}\}, \{x_{\text{dist}}\})  &\propto \mP(\{z_{\text{obs}}\}, \{x_{\text{dist}}\}|H_0) \mP(H_0) \nonumber \\
     &= \mP(H_0)\prod_{i=1}^{N_{\text{events}}}\mP(z^i_{\text{obs}}, x^i_{\text{dist}}|H_0).
\end{align}
In order to go further, we simplify the likelihood, $\mP(z^i_{\text{obs}}, x^i_{\text{dist}}|H_0)$, by expanding it in terms of the intermediate distance variable, $d$, and the velocity field, $\mvec{v}$, used to correct the observed redshifts.
\begin{equation}
    \mP(z^i_{\text{obs}}, x^i_{\text{dist}}|H_0) = \int \diffop d\; \mP(z^i_{\text{obs}}|H_0, \mvec{v}, d)\mP(x^i_{\text{dist}}|d)\mP(d).
\end{equation}
Finally, we have to model the likelihood for the redshift, $\mP(z^i_{\text{obs}}|H_0, \mvec{v}, d)$. 
\begin{equation}
    \mP(z^i_{\text{obs}}|H_0, \mvec{v}, d) = \frac{1}{\sqrt{2\pi\sigma^2_v}}\exp\bigg[ -\frac{(cz^i_{\text{obs}} -cz_{\text{pred}})^2}{2\sigma^2_v}\bigg],
\end{equation}
where, $cz_{\text{pred}} \equiv cz_{\text{pred}}(H_0, d, \mvec{v})$ and $\sigma_v$ is the typical uncertainty in the predicted peculiar velocity.

\subsection{Peculiar velocity of NGC 4993}\label{ssec:ngc4993}

The exact peculiar velocity of NGC4993 has been the discussion of many recent works \citep[e.g., ][]{lahav_paper, howlett_davis, borg_velocity_corrections}. NGC4993 was the host galaxy for the binary neutron star event, GW170817 \citep{gw170817} discovered by LIGO. The distance measurement from the gravitational wave event can be used to put constraint on the value of $H_0$. Because it is relatively nearby, the contribution to the redshift from the peculiar velocity is substantial ($\gtrsim~10\%$) and needs to be corrected. 

The original estimate of $H_0$ from GW170817 \citep{ligo_H0}, used adaptive kernel smoothing on the 6dF peculiar velocity sample to predict the velocity of NGC4993. It was noted that reconstruction based method with 2M++ also gives a similar velocity estimate. \citet{howlett_davis} tested the dependence of the peculiar velocity predictions on different assumptions, such as group assignment and different peculiar velocity catalogues for kernel smoothing. It was demonstrated that different assumptions leads to different estimates for the peculiar velocity, thus leading to a larger uncertainty in the measured value of $H_0$. \citet{borg_velocity_corrections} used the forward-modelled reconstruction framework, {\sc borg} \citep{borg_original, borg_pm}, to predict the velocity field with the 2M++ catalogue, finding a velocity estimate of $330 \pm 130$ km s$^{-1}$ for NGC4993.

\begin{figure*}
    \centering
    \includegraphics[width=\linewidth]{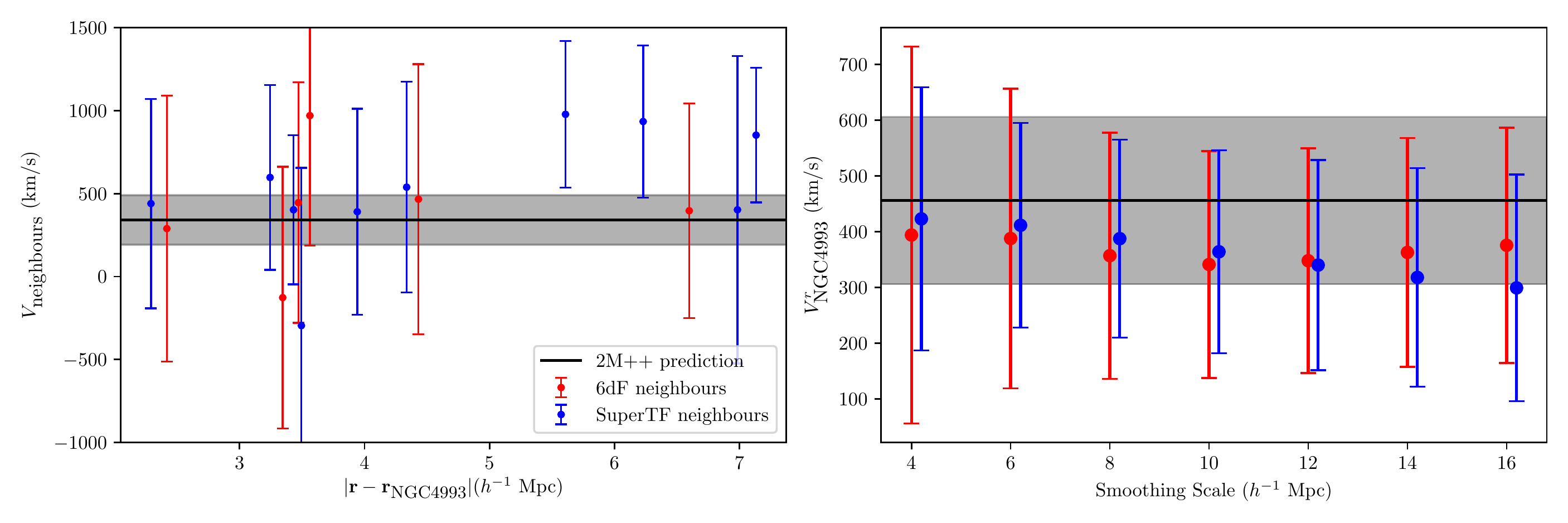}
    \caption{Peculiar velocity estimate of NGC4993. {\it (Left)} The measured peculiar velocity of the galaxies within 1 kernel width (for a fiducial smoothing length of $8~h^{-1}$ Mpc) of NGC4993 from the 6dF (red markers) and the SuperTF (blue markers) peculiar velocity catalogues. There are a total of $6$ neighbours within 1 kernel width for the 6dF catalogue and $10$ neighbors within a kernel width for the SuperTF catalogue. {\it (Right)} Predicted peculiar velocity for different models as a function of the smoothing scale. The black horizontal line is the prediction from 2M++ reconstruction and the grey region is the associated error, assuming $\sigma_v = 150$ km/s. The red and the blue markers are the estimates using the adaptive kernel smoothing on 6dF and SuperTF catalogues respectively for different smoothing scales.}
    \label{fig:ngc_4993}
\end{figure*}

In this section, we check the predictions for the peculiar velocity of NGC4993 using different choices of the peculiar velocity models considered in this work. \citet{lahav_paper} noted that the estimates of the peculiar velocity for NGC4993 using the kernel smoothing technique depends strongly on the choice of kernel width and used a Bayesian model to account for the uncertainty due to the choice of the kernel width. As we already point in Section \ref{sec:scaling}, predictions from a smoothed velocity field need to be scaled up in order to obtain unbiased estimates for the velocity. We show the dependence of the predicted peculiar velocity of NGC4993 on the smoothing length in right panel of Figure \ref{fig:ngc_4993}. As can be seen, after scaling the velocity by the required factor, there is no longer a strong dependence on the smoothing scale. The predictions from 2M++ reconstruction and adaptive kernel smoothing give consistent results. Since our methods of estimating the peculiar velocity smooths the velocity field at a scale much larger than that of individual galaxies, it is useful to predict the velocities of groups of galaxies. Averaging the redshifts of galaxies in a group suppresses the non-linear velocity contributions. In order to correct for the grouping, we identify the groups from the 2M++ catalogue. For NGC4993, the average group redshift is $cz = 3339$ km s$^{-1}$, while that of the galaxy is $cz = 3216$ km/s. Using a fiducial smoothing length of $8~h^{-1}$ Mpc, the adaptive kernel smoothing technique predicts a velocity of $v_r = 357 \pm 219$ km/s for the 6dF catalogue and $v_r = 388 \pm 162$ km/s for the SuperTF catalogue. The uncertainty for the two fields are obtained by adding the measurement error from the kernel smoothing added in quadrature to $\sigma_v= 150$ km/s. The justification for using this uncertainty is as follows. The uncertainty in the peculiar velocity estimated using the kernel smoothing has two contributions: {\it i)} The measurement uncertainty for the peculiar velocity data, and, {\it ii)} The scatter of the peculiar velocity of the galaxies from which the kernel-smoothed velocity is estimated. The first part can be estimated from the measurement errors on the peculiar velocity estimates. For the second part, we assume a value, $\sigma_v = 150$ km/s. Since the 2M++ reconstruction gives the velocity in the real space coordinates, we use an iterative method to estimate the velocity along the line-of-sight of the galaxy. For NGC4993, we find the 2M++ predictions to be, $v_r = 456 \pm 150$ km/s. Therefore, we find consistent predictions from the 2M++ reconstruction and the kernel smoothing methods.  

\begin{figure}
    \centering
    \includegraphics[width=\linewidth]{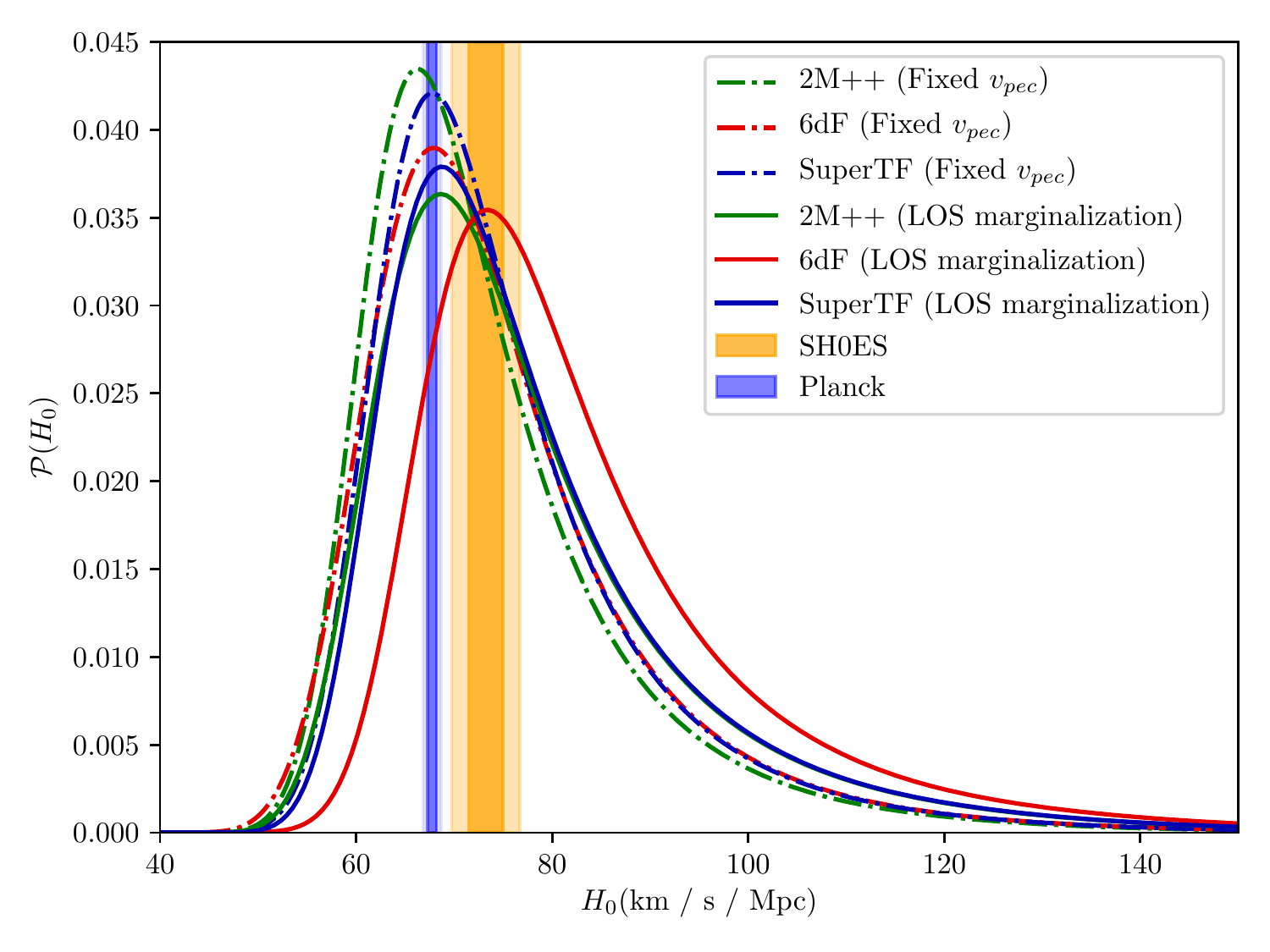}
    \caption{$H_0$ posterior for NGC4993 considering the distances measured from GW170817. We show the effect of marginalizing the peculiar velocity correction along the line-of-sight. The solid lines uses the likelihood of section \ref{ssec:H0_lkl}, while the dash-dotted lines are obtained by assuming a fixed peculiar velocity estimate using the likelihood of \citet{ligo_H0}. The peculiar velocity is corrected using the 2M++ reconstruction ({\it red curves}), 6dF adaptive kernel smoothing ({\it green curves}) and SuperTF adaptive kernel smoothing ({\it blue curves}). The Planck and the SH0ES confidence intervals are shown with blue and orange vertically-shaded regions respectively.}
    \label{fig:H0_posterior_ngc4993}
\end{figure}

We also used the likelihood described in section \ref{ssec:H0_lkl} to derive the constraint on $H_0$ from the gravitational wave distance. For NGC 4993, the luminosity distance can be measured from the gravitational wave event GW170817 \footnote{The posterior samples are publicly available at \url{https://dcc.ligo.org/LIGO-P1800061/public}}. A volumetric prior, $\mP_{\text{prior}}(d) \propto d^2$, was used inferring the distance posterior. Therefore, we also used a volumetric prior for inferring $H_0$ with the GW data. We compare the results from the likelihood for measuring $H_0$ that was used in \citet{ligo_H0}, where a fixed velocity estimate was used. The inferred value of $H_0$ for the different peculiar velocity fields and the two likelihoods are given in Table \ref{tbl:std_siren_H0}. The $H_0$ posterior using the GW distance is shown in Figure \ref{fig:H0_posterior_ngc4993}. We show the effect of using the different peculiar velocity models considered in this work. The table and the figure shows that marginalizing the peculiar velocity correction along the line-of-sight inflates the error bar by $\sim 20\%$. 

\begin{table}
  \centering
  \caption{Inferred value of $H_0$ from GW170817 for different treatments of peculiar velocities. Results are reported as the median with $1\sigma$ confidence interval (16th to 84th percentile).}
  {\renewcommand{\arraystretch}{1.2}
  \begin{tabular}{l l c}
  \hline

      $H_0$ Likelihood & Velocity model & $H_0$ (\kmsMpc)\\
    \hline
    \citet{ligo_H0} & 2M++ & $70.5^{+14.2}_{-8.8}$ \\
   (Fixed $v_{\text{pec}}$) & 6dF& $72.0^{+14.9}_{-9.2}$ \\
    & SuperTF & $72.1^{+14.5}_{-8.5}$ \\
    \hline
     Section \ref{ssec:H0_lkl} & 2M++ & $74.3^{+16.9}_{-9.6}$ \\
    (Line-of-sight& 6dF& $79.1^{+18.1}_{-9.6}$ \\
    marginalization) & SuperTF & $74.3^{+17.5}_{-9.0}$ \\
    \hline
  \end{tabular}}
  \label{tbl:std_siren_H0}
\end{table}

\begin{figure}
    \centering
    \includegraphics[width=\linewidth]{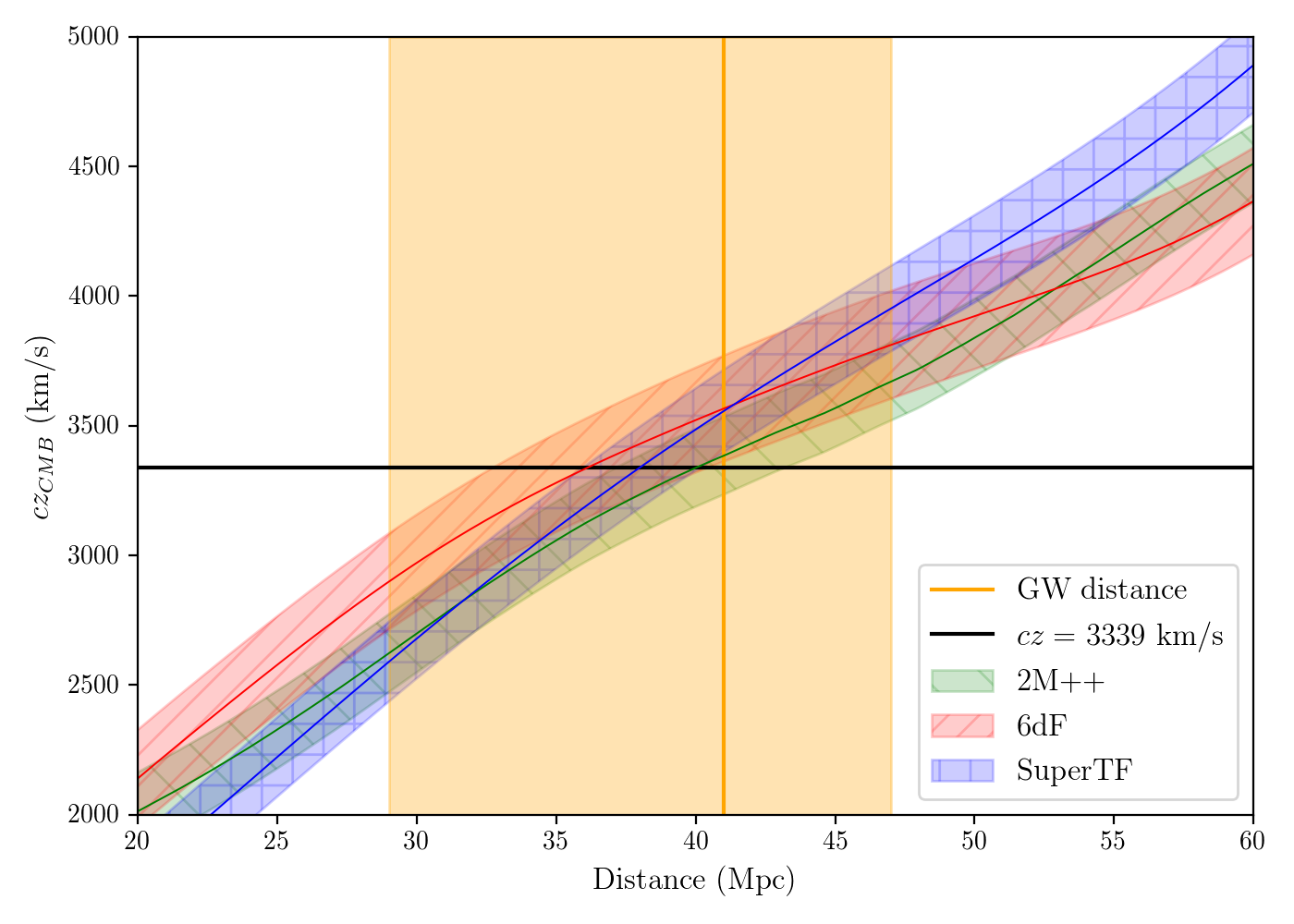}
    \caption{Predicted $cz_{\text{CMB}}$ as a function of distance along the line of sight to NGC 4993. We use $h=0.72$ to convert the distance in $h^{-1}$ Mpc to absolute distance. The shaded regions for the 2M++, 6dF and the SuperTF radial velocities (shown in green, red and blue respectively) signify the $1\sigma$ errors associated with the velocity estimates. The CMB frame redshift for the NGC 4993 group ($3339$ km/s) is  shown with a solid black line. The orange vertical band shows the distance measurement to NGC 4993 from the gravitational wave detection.
    }
    \label{fig:ngc4993_los_cz}
\end{figure}
\subsection{Peculiar velocity correction for megamasers}

Water megamasers in the active galactic nuclei of galaxies provide a completely geometric method to measure distances without the need for intermediate calibration. In \citet{MasersH0}, the Megamaser Cosmology Project (MCP) measured the value of $H_0$ from the distance measurement to $6$ such megamasers. It was found that the inferred value of $H_0$ depended strongly on the treatment of peculiar velocities.

In this section, we reanalyze the distance data from the megamasers to infer the constraints on $H_0$, checking different assumptions for peculiar velocity correction. Our treatment is different from \citet{MasersH0} in four ways:
\begin{itemize}
    \item[{\it i)}] We used the likelihood of section \ref{ssec:H0_lkl} to marginalize over the line-of-sight peculiar velocity as opposed to using a point estimate for the peculiar velocity.
    \item[{\it ii)}] We use a volumetric prior for the distances.
    \item[{\it iii)}] We use the group-corrected redshift instead of the individual redshift for each galaxy.
    \item[{\it iv)}] We use two different values of $\sigma_v$ to check the robustness of the result to the choice of this parameter.
\end{itemize}

We discuss the effect of each of these modelling assumptions on the inferred value of $H_0$ in the following. We use the peculiar velocity fields described in the previous sections to correct for peculiar velocities. Since all $6$ megamasers are located in the northern sky, we cannot use the 6dF velocity field for our purpose. We used both the 2M++ and the SuperTF velocity fields for our treatment of the peculiar velocity.

\underline{\textit{Marginalizing the line-of-sight peculiar velocity}}: We use the probability model described in section \ref{ssec:H0_lkl}, where we marginalize over the line-of-sight peculiar velocity, as opposed to a point estimate that is usually used. We compared the effect of not marginalizing the line-of-sight peculiar velocity. When using the simple point estimate of the peculiar velocity, we find that the $H_0$ posterior shifts by $\sim +1.5$ \kmsMpc. This is a non-negligible effect on the inferred value of $H_0$. Such a shift is not present when using the SuperTF velocities, suggesting that the effect of marginalizing over the line-of-sight peculiar velocity may be non-trivial. In addition to using a fixed peculiar velocity, if we use a uniform prior on the distances, we get $H_0 = 71.5 \pm 2.7$ \kmsMpc, which is consistent with the values obtained by \citet{MasersH0} for the 2M++ velocity field. 

\underline{\textit{Distance priors}}: We use a volumetric prior on the distances. It has been noted that assuming a wrong distribution biases the distance measurement, an effect usually called \citet{Mal20} bias, although the first derivation of this effect is in \citet{eddington14}. The fractional bias for a lognormal uncertainty is given as, $3\Delta^2$, where $\Delta$ is the fractional distance uncertainty \citep{LyndenBell88}. Therefore assuming a uniform prior on the distances, instead of a volumetric prior, will bias the distance measurements to a lower value, leading to an inferred value of $H_0$ that is systematically biased high. For the megamasers, the difference in the prior distribution biases the value of $H_0$ by $\sim 1$ \kmsMpc\ (see Table \ref{tbl:mega_maser_H0}).

\begin{table}
  \centering
  \caption{Inferred value of $H_0$ from megamasers for different treatment of peculiar velocities. Results are reported as the median with $1\sigma$ confidence interval (16th to 84th percentile).}
  {\renewcommand{\arraystretch}{1.2}
  \begin{tabular}{l l c}
  \hline
      Peculiar velocity & Model Assumption & $H_0$ (km/s/Mpc)\\
    \hline
   2M++ & \textbf{Fiducial} & $\boldsymbol{69.0^{+2.9}_{-2.8}}$ \\
    & Fixed $v_{\text{pec}}$ & $70.5^{+2.6}_{-2.8}$ \\
    & Uniform distance prior & $70.1 \pm 2.9$ \\
   & No group redshift correction & $68.6^{+2.9}_{-2.8}$ \\
   & $\sigma_v = 200$ km/s & $69.4^{+3.1}_{-3.0}$ \\
   & Fixed $v_{\text{pec}}$ and uniform prior & $71.5 \pm 2.7$\\
   & \citet{MasersH0} 2M++ fit& $71.8\pm2.7$\\
    \hline
    SuperTF & \textbf{Fiducial} & $\boldsymbol{72.6^{+7.1}_{-6.3}}$ \\
    & Fixed $v_{\text{pec}}$ & $72.0^{+7.0}_{-6.5}$ \\
    & Uniform distance prior & $74.4^{+7.3}_{-6.5}$ \\
   & No group redshift correction & $72.2^{+7.0}_{-6.1}$ \\
    \hline
  \end{tabular}}
  \label{tbl:mega_maser_H0}
\end{table}

\begin{figure}
    \centering
    \includegraphics[width=\linewidth]{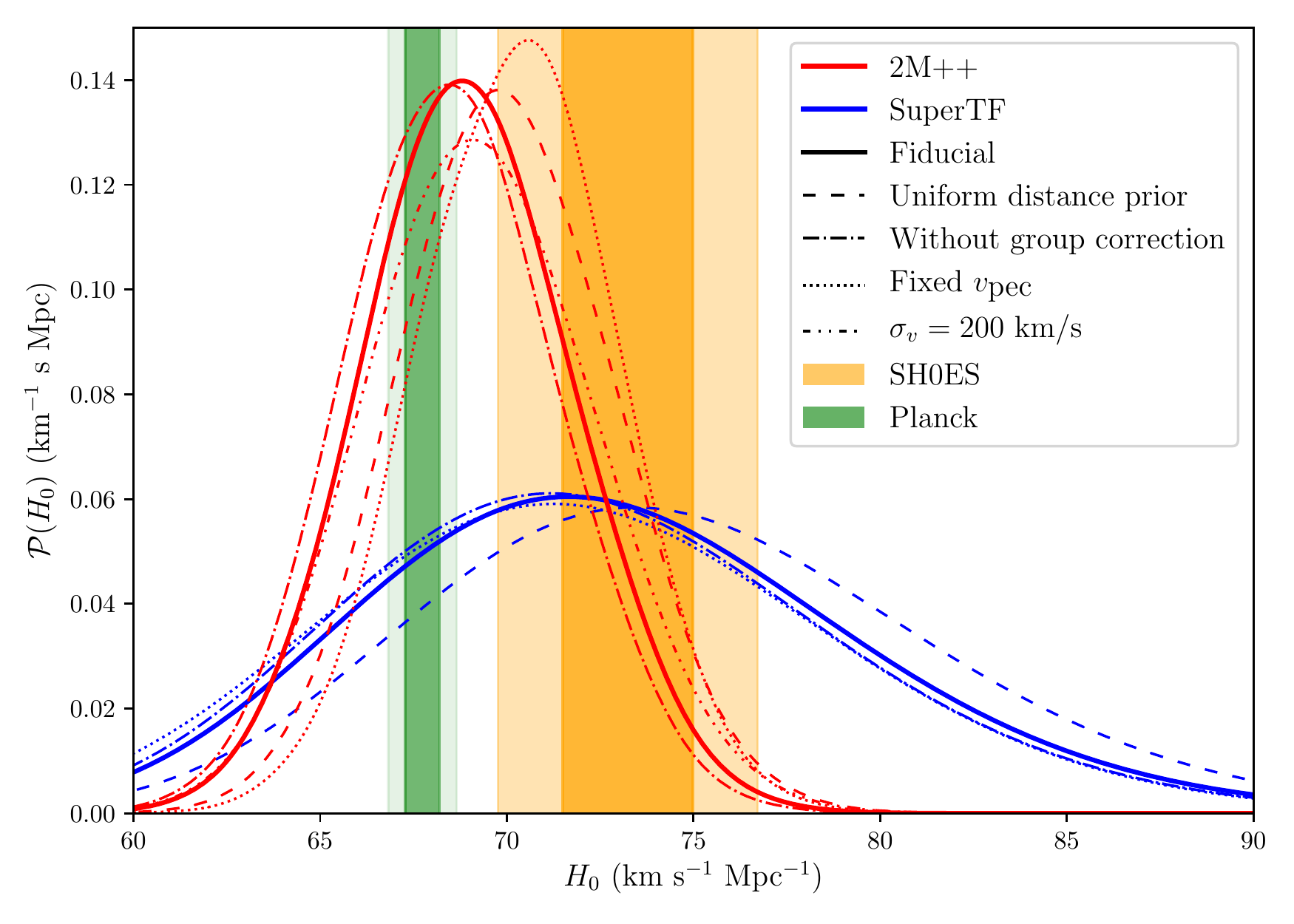}
    \caption{$H_0$ posteriors from the megamaser distances for different peculiar velocity treatments. The red curves are obtained using the 2M++ velocity field, while blue curves are obtained using the SuperTF velocity field. Different line styles corresponds to different model assumptions as indicated in the plot. The Planck and SH0ES confidence intervals are shown with green and orange vertical bands.}
    \label{fig:mega_maser_H0}
\end{figure}

\begin{figure}
    \centering
    \includegraphics[width=\linewidth]{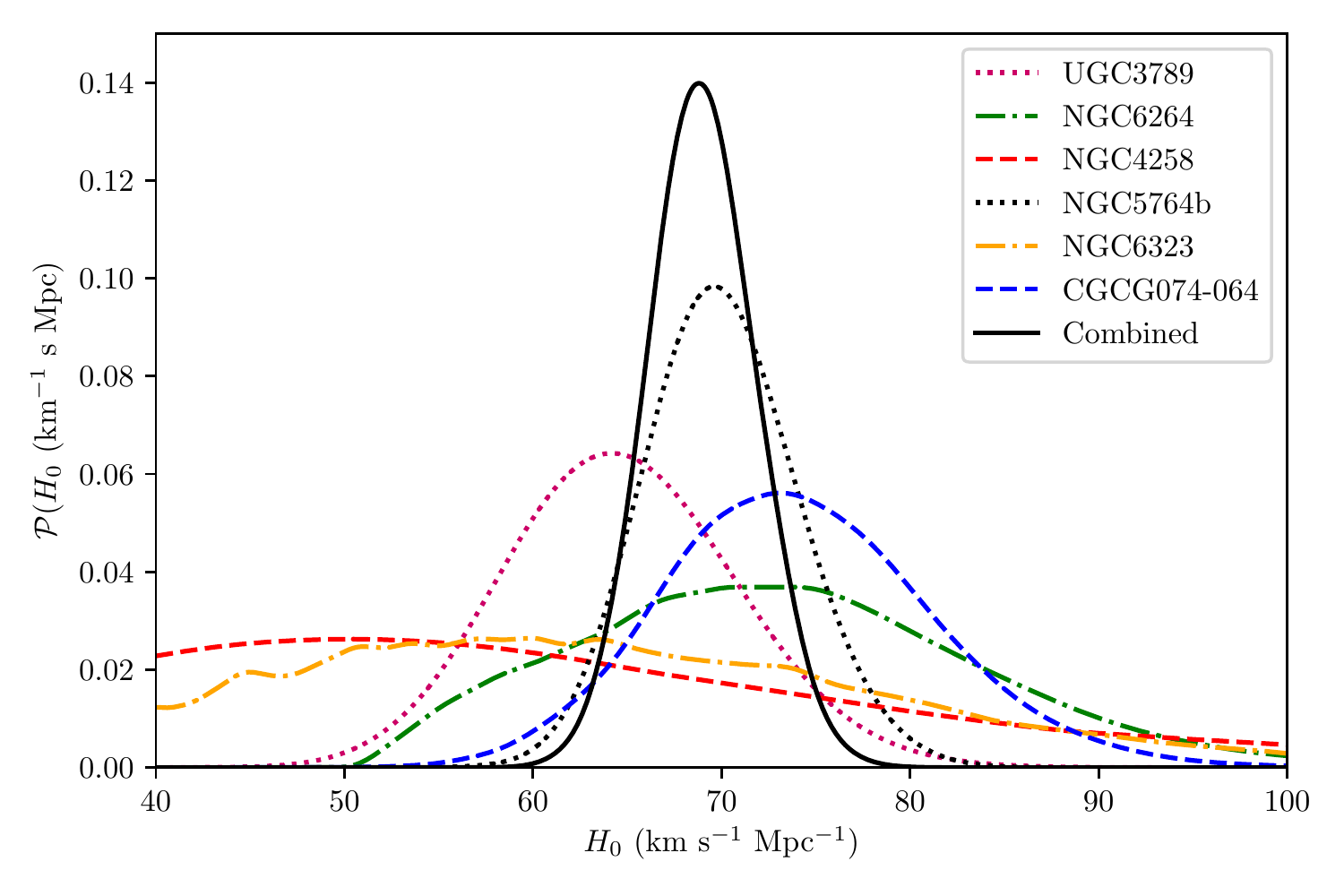}
    \caption{The $H_0$ posterior for each of the individual megamasers for our fiducial model assumptions with the 2M++ velocity field. NGC5764b provides the tightest constraint on $H_0$ among the $6$ megamasers.}
    \label{fig:maser_individual_posteriors}
\end{figure}

\begin{figure*}
    \centering
    \includegraphics[width=\linewidth]{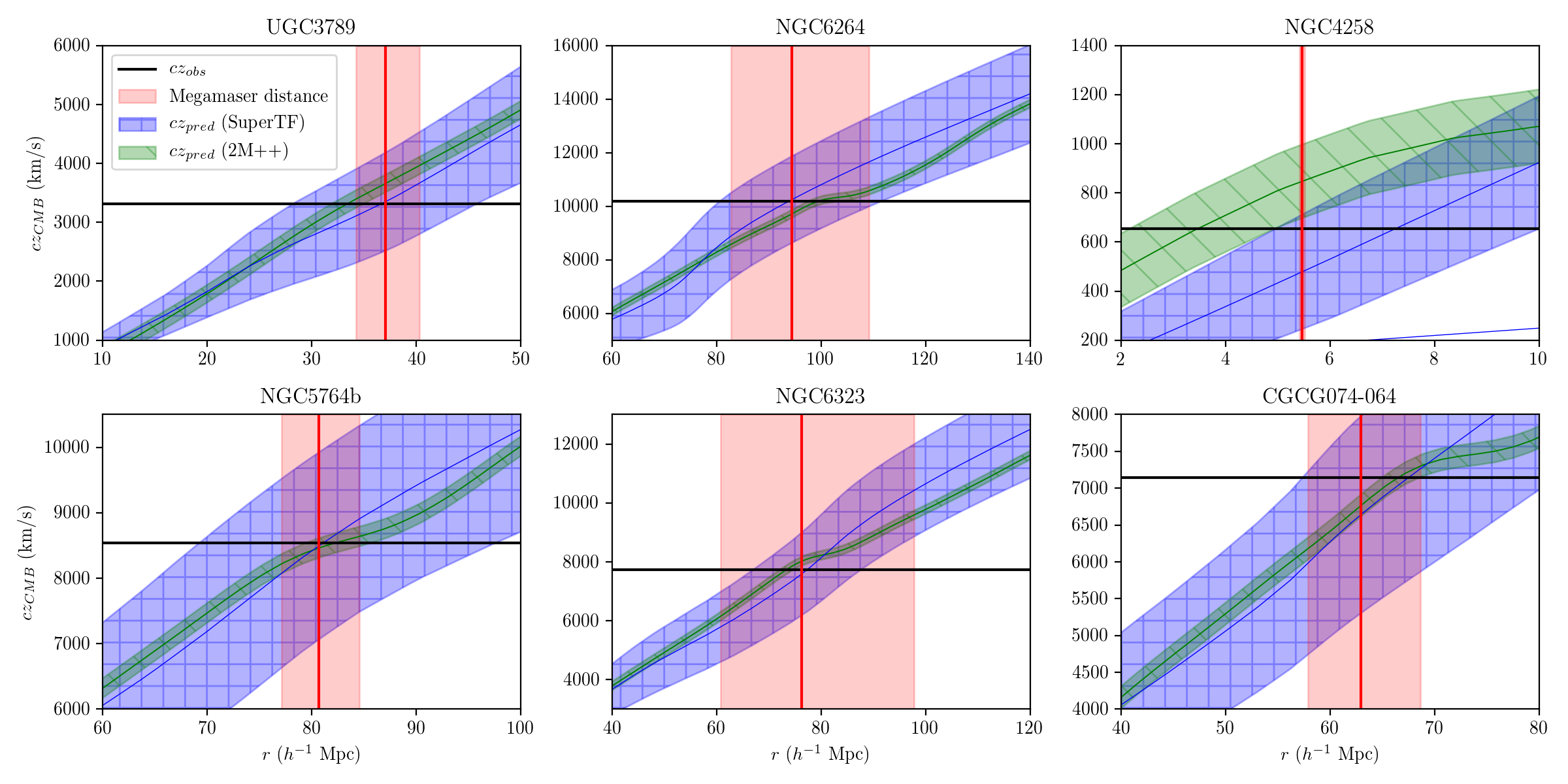}
    \caption{The predicted CMB frame redshifts along the line-of-sights of the 6 megamasers. The red shaded region shows the distance uncertainty as measured by \citet{MasersH0}. We use $h = 0.72$ to convert the angular diameter distances to $r$ (measured in $h^{-1}$ Mpc). The black horizontal line is the observed CMB frame redshift, with the group corrections. The green and the blue shaded region is the uncertainty in the predictions of $cz$ from 2M++ and SuperTF fields respectively.}
    \label{fig:megamaser_los_cz}
\end{figure*}

\underline{\textit{Group redshift corrections}}: 
As in the previous section, we use the group mean redshifts for the megamaser galaxies to suppress the non-linear velocity contributions. Except for NGC 4258, we identify the groups from the 2M++ catalogue. In \citet{Kourkchi17}, it was noted that the NGC 4258 group is located directly in the foreground of the NGC 4217 group. This can lead to mistaken identification of the two groups into the same group. In the group catalogue of 2M++, the two groups are identified into a single group. Therefore, we use the group information for NGC 4258 from \citet{Kourkchi17}. We note that for the MCP megamaser sample, using the group redshift has a small effect, leading to a value $H_0$ that is $0.4$ \kmsMpc{} higher than when using the galaxy redshifts.

\underline{\textit{Value of $\sigma_v$}}: We also checked the effect of using a different value of $\sigma_v$ for the peculiar velocity error. The $\sigma_v = 150$ km/s uncertainty was obtained by calibrating our reconstruction with simulations. If for some reason, the peculiar velocity error is underestimated, the value of $\sigma_v$ may be higher. We therefore test the effect of adopting a larger uncertainty, $\sigma_v = 200$ km/s for our measurements. The effect of different values of $\sigma_v$ in non-trivial. As expected, the uncertainty increases for the higher $\sigma_v$ value. The mean inferred value $H_0$ is also $0.4$ \kmsMpc\ higher than for the fiducial value of $\sigma_v$ .

We present our results in Table \ref{tbl:mega_maser_H0}, where the inferred value of $H_0$ for different model assumptions are reported for the 2M++ and the SuperTF peculiar velocity fields. For the SuperTF velocity field, we add in quadrature to $\sigma_v = 150$ km/s, the measurement uncertainty in the kernel-smoothed velocity field. As can be seen in Figure \ref{fig:z_dist}, the distribution of SuperTF galaxies at high redshift is sparse. Furthermore, as discussed in section \ref{ssec:compare_aks_rec}, at higher redshift, the peculiar velocity uncertainties are also larger. Hence, the velocity corrections for the megamasers obtained using the SuperTF field is a very noisy, leading to the much larger uncertainty in $H_0$. In the `fiducial model', we marginalize over the line-of-sight peculiar velocity, assume a volumetric prior for the distance, and the group corrected redshifts are used. For 2M++, we assume $\sigma_v = 150$ km s$^{-1}$ for the fiducial model. The posteriors derived for the different model assumptions are shown in Figure \ref{fig:mega_maser_H0}. As can be seen from the figure, we find that using the SuperTF velocity field yields a much larger uncertainty in $H_0$ compared to the 2M++ velocity field. Of the different assumptions that we checked, we highlight the importance of using a volumetric prior and marginalizing the line-of-sight peculiar velocity. For the 6 megamasers, a combination of these two effects shifts the $H_0$ posterior by $\sim 1\sigma$. In Figure \ref{fig:maser_individual_posteriors}, we show the $H_0$ posterior from each of the 6 individual megamasers. As can be seen from the figure, NGC5764b provides the strongest constraints among the 6 megamasers. We show the predicted redshift along the line-of-sight of the megamasers from the 2M++ and the SuperTF velocity fields in Figure \ref{fig:megamaser_los_cz}. The inferred value of $H_0$ is broadly consistent with both the value of $H_0$ as inferred by Planck and the SH0ES collaboration.


\section{Distance and peculiar velocity of NGC 1052-DF2}\label{sec:ngc1052}

\begin{figure*}
    \centering
    \includegraphics[width=\linewidth]{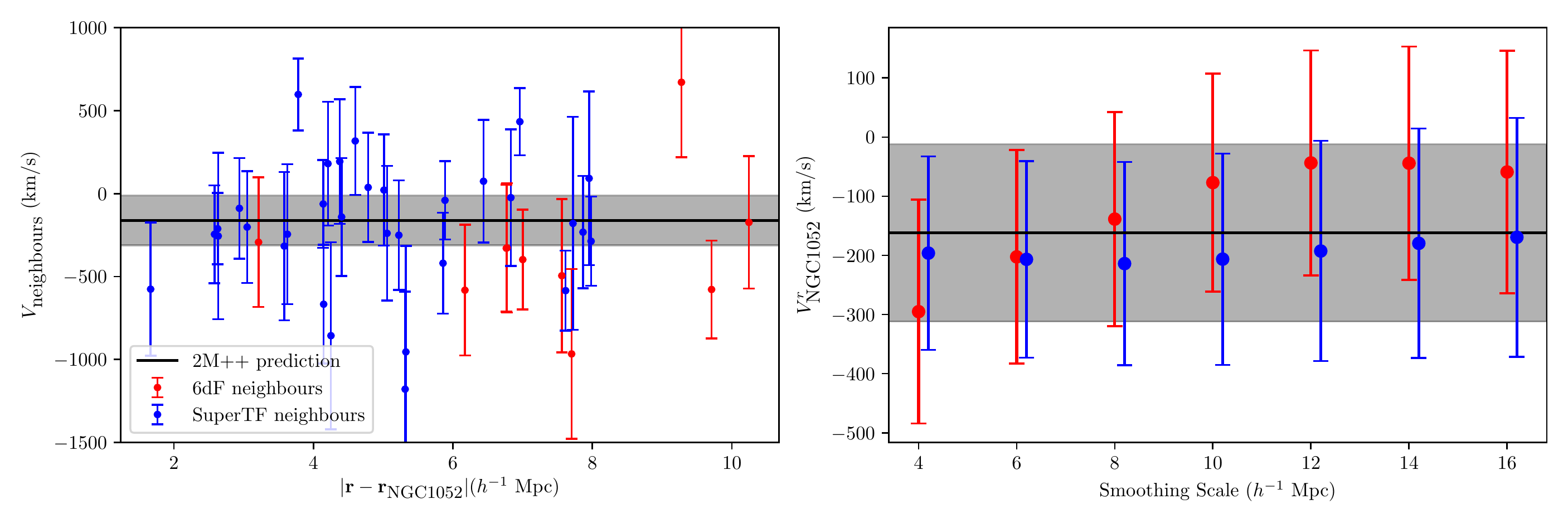}
    \caption{Same as Figure \ref{fig:ngc_4993} but for NGC1052-DF2. }
    \label{fig:ngc1052_vel}
\end{figure*}

In an interesting result, \citet{ngc1052_discovery} discovered that a galaxy, NGC1052-DF2, contains little or no dark matter. This challenges the conventional wisdom about galaxy formation and shows that at least some galaxies may have baryonic component without any dark matter in it. However, their result was contested by \citet{ngc1052_distance1} where, using Tip of Red Giant Branch (TRGB) and other distance measurements, the authors calculated a distance of $\sim 13$ Mpc to the galaxy as opposed to the earlier estimates of $\sim 19$ Mpc as derived from surface brightness fluctuations in \citet{ngc1052_discovery}. An analysis adopting the shorter distance to NGC1052-DF2 results in total-mass-to-stellar mass ratio, $M_{\text{halo}}/M_{\star} > 20$ as opposed to the value of order unity derived in \citet{ngc1052_discovery}.

However, the shorter (13 Mpc) distance implies a radial peculiar velocity of $640\pm25$ km/s in the CMB frame for NGC 1052-DF2. While \citet{ngc1052_distance1} claim that the spread of peculiar velocities in this region is high, we argue that most of this spread is due to uncertainty in the distance indicator and does not reflect the real velocity noise on the underlying flow field.  For example, for the TF and FP methods, the scatter in the distance indicator is $\sim 20\%$, which translates to a scatter of $\sim 350$ km/s in the peculiar velocity from galaxy to galaxy at the distance of NGC 1052. In this section, we infer the peculiar velocity of the galaxy by the two methods previously described. The result of this analysis is shown in Figure \ref{fig:ngc1052_vel}. The mean redshift of the 9 members of the NGC 1052 group (in which NGC1052-DF2 is assumed to reside) as identified in the 2M++ catalogue is $cz_{\text{CMB}} = 1256$ km/s. This is in agreement with \citet{Kourkchi17}, who find a mean CMB redshift of 1252 km/s from 16 group members.  Note that this is considerably lower than the CMB redshift of NGC 1052-DF2 itself (1587 km/s). Using the 2M++ reconstruction, the peculiar velocity for NGC 1052 group is $v_r = -162 \pm 150$ km s$^{-1}$. Assuming 
$H_0 = 72$ \kmsMpc, this peculiar velocity implies a distance of $19.7\pm2.1$ Mpc.
The distance estimates are consistent with the Fundamental Plane distance estimate for NGC1052, $d = 19.4 \pm 2.4$ Mpc \citep{SBFIV} and the SBF distances derived for NGC1052-DF2 itself, $19\pm1.7$ Mpc and $20.4\pm2.0$ Mpc by \citet{ngc1052_discovery} and \citet{BlaCan18}, respectively. We find consistent results from the kernel smoothing approaches: with a fiducial smoothing radius of $8~h^{-1}$ Mpc, with the 6dF velocity data there are a total of 9 6dF velocities within 1 kernel length of NGC1052-DF2 and these yield a kernel-smoothed mean velocity of $v_r = -124 \pm 165$ km s$^{-1}$. With the SuperTF data, the kernel smoothed mean velocity is $v_r = -191 \pm 155$ km s$^{-1}$ from 32 SuperTF velocities within 1 kernel length of NGC 1052-DF2. 

\begin{figure}
    \centering
    \includegraphics[width=\linewidth]{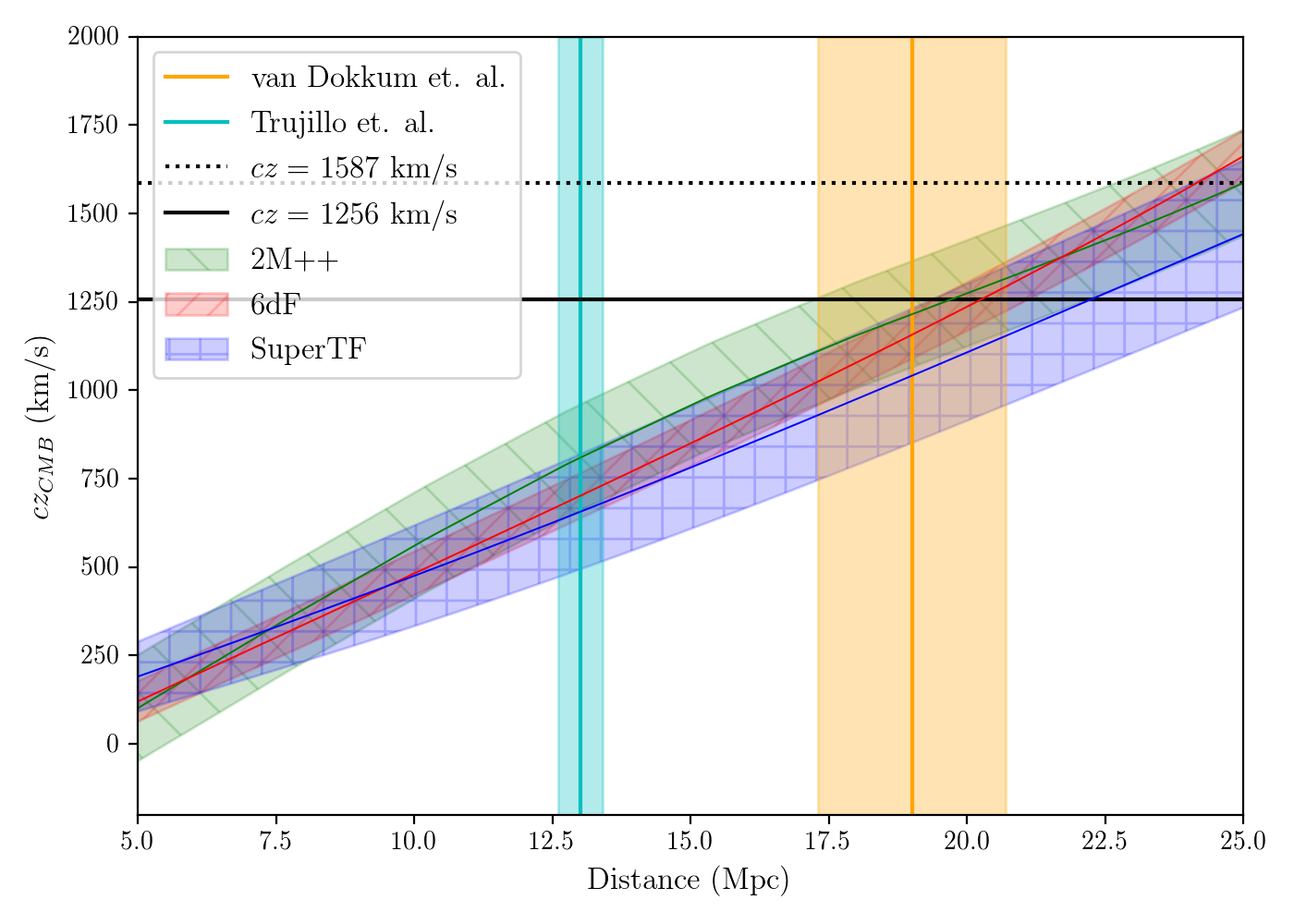}
    \caption{Predicted $cz_{\text{CMB}}$ as a function of distance along the line of sight to NGC 1052-DF2. We use $h=0.72$ to convert the distance in $h^{-1}$ Mpc to absolute distance. The shaded regions for the 2M++, 6dF and the SuperTF radial velocities (shown in green, red and blue respectively) signify the $1\sigma$ errors associated with the velocity estimates. The observed redshift for NGC 1052-DF2 is shown with a dotted black line. The redshift of the NGC 1052 group is shown with a solid black line. The orange vertical band shows the surface brightness fluctuation distance to NGC 1052 from \citet{ngc1052_discovery}. The vertical cyan band shows the distance for NGC 1052-DF2 advocated by \citet{ngc1052_distance1}.}
    \label{fig:ngc1052_los_cz}
\end{figure}

We also consider the possibility that NGC 1052-DF2 may be in the foreground of the NGC 1052 group with a distance of $\sim 13$ Mpc. In this case, if it is isolated, then the peculiar velocity needed to explain its redshift is $\sim 600$ km s$^{-1}$. In Figure \ref{fig:ngc1052_los_cz}, we plot the predicted redshift along the line-of-sight of NGC1052-DF2 for the different peculiar velocity models. As can be seen from the Figure, there are no locations in the foreground of the NGC 1052 group with such high outward radial peculiar velocity.

More recently, \citet{MonTru19} have suggested that NGC1052-DF2 is a member of a foreground group dominated by the spirals NGC1042 and NGC1035 at a distance of $\sim 13.5$ Mpc. The mean CMB velocity of these two galaxies is $1184$ km/s, which would mean that NGC 1052-DF2 would have a peculiar velocity of $403$ km/s with respect to this poor group: unlikely, but perhaps not impossible if NGC1052-DF2 is close to the bottom of the group's potential well and is falling in directly along the line-of-sight. However the group itself, if at a distance of 13.5 Mpc, would have a CMB frame peculiar velocity (assuming $H_0 = 72$ \kmsMpc) of $+212$ km/s, where the 2M++ predicted peculiar velocity is $-139\pm 150$ km/s, a discrepancy significant at the $\sim 2.3 \sigma$ level. Turning the problem around, to be consistent with the 2M++ predictions, the group's distance must be greater than $14.2$ Mpc at the 95\% confidence level. Therefore, we conclude that a short distance is not a likely explanation for the anomalously low dark matter mass fraction of NGC 1052-DF2.


\section{Summary}\label{sec:summary}

Unbiased estimates of peculiar velocity are essential for different applications in cosmology and galaxy formation. There are different methods of estimating the peculiar velocities of galaxies, e.g, using reconstruction of the density field in the local Universe or by kernel smoothing the peculiar velocity data. In this work, we compared the performance of different peculiar velocity models of the local Universe. The first model we studied is the reconstructed velocity field from the 2M++ redshift compilation. The others are based on an adaptive kernel smoothing technique, which we apply on the 6dF peculiar velocity data and SuperTF, a compilation of the Tully-Fisher peculiar velocity data from SFI++ and 2MTF. We highlight that, when using a smoothed velocity field, we need to rescale the predictions by a scaling factor to get unbiased estimate of the peculiar velocity. We compared the peculiar velocity predictions to a few test data sets using a simple comparison of the mean squared error and a forward likelihood method. We find that the 2M++ reconstruction performs better than both the kernel smoothed peculiar velocity data for all the peculiar velocity test data sets across all range of redshifts. 

We also compared the peculiar velocity estimates from these different methods for a few specific galaxies. First, we investigated the implications of our peculiar velocity fields for the measurement of $H_0$ from standard sirens and megamasers. In doing so, we introduced a probabilistic framework where we marginalize over the line-of-sight peculiar velocity to accurately capture the effect of peculiar velocity corrections. NGC4993 was the host galaxy for the first binary neutron star event detected by LIGO. Because of it nearby location, accurate peculiar velocity is required to correct the redshift of the galaxy in order to obtain the measurement of the Hubble constant. The different models considered in this work give remarkably consistent peculiar velocity estimates for the galaxy. For NGC4993, we notice that marginalizing over the line-of-sight peculiar velocity inflates the uncertainty on $H_0$ by a factor of $\sim 1.5$. Another distance indicator that does not rely on intermediate distance calibrator is megamaser. We also checked the effects of different assumptions about the peculiar velocity correction on the inferred value of $H_0$ from the megamasers. We highlight two key factors that can significantly bias the inferred value of $H_0$ from the megamasers: {\it i)} using the wrong prior for the distances, and {\it ii)} not marginalizing over the line-of-sight peculiar velocity. With our fiducial model assumptions with the 2M++ velocity field, we find $H_0 = 69^{+2.9}_{-2.8}$ \kmsMpc, which is $\sim 1.5\sigma$ lower than the value obtained by the SH0ES collaboration from Type Ia supernovae. Finally, we also investigate the peculiar velocity of NGC 1052-DF2, which is an ultra-diffuse galaxy that has been claimed to be almost free of dark matter. This result has been contested with the claim that a shorter distance to the galaxy solves the anomalous stellar mass fraction. However, we find that this claim is not supported by the models of peculiar velocity that we use in our study.

\section*{Acknowledgements}

We thank Ofer Lahav and Cullan Howlett for useful discussions that stimulated this project and for valuable suggestions, and Dom Pesce and Jim Braatz for providing the distance measurement samples from the Megamaser Cosmology Project (MCP).  This work has been done as part of the activities of the Domaine d'Int\'{e}r\^{e}t Majeur (DIM) ``Astrophysique et Conditions d'Apparition de la Vie'' (ACAV), and received financial support from R\'{e}gion Ile-de-France. GL acknowledges financial support from the ANR BIG4, under reference ANR-16-CE23-0002.  MH acknowledges the support of an NSERC Discovery Grant. GL thanks the University of Helsinki for hospitality. This work is done within the Aquila Consortium\footnote{\url{https://www.aquila-consortium.org/}}.

\section*{Data availability statement}
The SFI++, 2MTF and the 6dF catalogues are publicly available with their respective publications as cited in section \ref{sec:pv_data}. The Second Amendment supernovae compilation is available with the supplementary data of \url{https://doi.org/10.1093/mnras/staa2485}. The 2M++ reconstruction used in this work is publicly available at \url{https://cosmicflows.iap.fr/}. Finally, the distance data from MCP were provided by Dom Pesce and Jim Braatz. 



\bibliographystyle{mnras}
\bibliography{peculiar_velocity_corrections} 

\begin{thebibliography}{}
\makeatletter
\relax
\def\mn@urlcharsother{\let\do\@makeother \do\$\do\&\do\#\do\^\do\_\do\%\do\~}
\def\mn@doi{\begingroup\mn@urlcharsother \@ifnextchar [ {\mn@doi@}
  {\mn@doi@[]}}
\def\mn@doi@[#1]#2{\def\@tempa{#1}\ifx\@tempa\@empty \href
  {http://dx.doi.org/#2} {doi:#2}\else \href {http://dx.doi.org/#2} {#1}\fi
  \endgroup}
\def\mn@eprint#1#2{\mn@eprint@#1:#2::\@nil}
\def\mn@eprint@arXiv#1{\href {http://arxiv.org/abs/#1} {{\tt arXiv:#1}}}
\def\mn@eprint@dblp#1{\href {http://dblp.uni-trier.de/rec/bibtex/#1.xml}
  {dblp:#1}}
\def\mn@eprint@#1:#2:#3:#4\@nil{\def\@tempa {#1}\def\@tempb {#2}\def\@tempc
  {#3}\ifx \@tempc \@empty \let \@tempc \@tempb \let \@tempb \@tempa \fi \ifx
  \@tempb \@empty \def\@tempb {arXiv}\fi \@ifundefined
  {mn@eprint@\@tempb}{\@tempb:\@tempc}{\expandafter \expandafter \csname
  mn@eprint@\@tempb\endcsname \expandafter{\@tempc}}}

\bibitem[\protect\citeauthoryear{{Abbott} et~al.}{{Abbott}
  et~al.}{2017a}]{gw170817}
{Abbott} B.~P.,  et~al., 2017a, \mn@doi [\prl]
  {10.1103/PhysRevLett.119.161101}, \href
  {https://ui.adsabs.harvard.edu/abs/2017PhRvL.119p1101A} {119, 161101}

\bibitem[\protect\citeauthoryear{{Abbott} et~al.}{{Abbott}
  et~al.}{2017b}]{ligo_H0}
{Abbott} B.~P.,  et~al., 2017b, \mn@doi [\nat] {10.1038/nature24471}, \href
  {https://ui.adsabs.harvard.edu/abs/2017Natur.551...85A} {551, 85}

\bibitem[\protect\citeauthoryear{{Adams} \& {Blake}}{{Adams} \&
  {Blake}}{2020}]{adams_and_blake_new}
{Adams} C.,  {Blake} C.,  2020, \mn@doi [\mnras] {10.1093/mnras/staa845}, \href
  {https://ui.adsabs.harvard.edu/abs/2020MNRAS.tmp.1095A} {}

\bibitem[\protect\citeauthoryear{{Behroozi}, {Wechsler}  \& {Wu}}{{Behroozi}
  et~al.}{2013}]{rockstar}
{Behroozi} P.~S.,  {Wechsler} R.~H.,   {Wu} H.-Y.,  2013, \mn@doi [\apj]
  {10.1088/0004-637X/762/2/109}, \href
  {https://ui.adsabs.harvard.edu/abs/2013ApJ...762..109B} {762, 109}

\bibitem[\protect\citeauthoryear{{Berlind}, {Narayanan}  \&
  {Weinberg}}{{Berlind} et~al.}{2000}]{unbiased_smoothing}
{Berlind} A.~A.,  {Narayanan} V.~K.,   {Weinberg} D.~H.,  2000, \mn@doi [\apj]
  {10.1086/309085}, \href
  {https://ui.adsabs.harvard.edu/abs/2000ApJ...537..537B} {537, 537}

\bibitem[\protect\citeauthoryear{{Blakeslee} \& {Cantiello}}{{Blakeslee} \&
  {Cantiello}}{2018}]{BlaCan18}
{Blakeslee} J.~P.,  {Cantiello} M.,  2018, \mn@doi [Research Notes of the
  American Astronomical Society] {10.3847/2515-5172/aad90e}, \href
  {https://ui.adsabs.harvard.edu/abs/2018RNAAS...2..146B} {2, 146}

\bibitem[\protect\citeauthoryear{{Boruah}, {Hudson}  \& {Lavaux}}{{Boruah}
  et~al.}{2019}]{sn_flows_paper}
{Boruah} S.~S.,  {Hudson} M.~J.,   {Lavaux} G.,  2019, arXiv e-prints, \href
  {https://ui.adsabs.harvard.edu/abs/2019arXiv191209383B} {p. arXiv:1912.09383}

\bibitem[\protect\citeauthoryear{{Campbell} et~al.,}{{Campbell}
  et~al.}{2014}]{6df_fp2}
{Campbell} L.~A.,  et~al., 2014, \mn@doi [\mnras] {10.1093/mnras/stu1198},
  \href {https://ui.adsabs.harvard.edu/abs/2014MNRAS.443.1231C} {443, 1231}

\bibitem[\protect\citeauthoryear{{Carrick}, {Turnbull}, {Lavaux}  \&
  {Hudson}}{{Carrick} et~al.}{2015}]{Carrick_et_al}
{Carrick} J.,  {Turnbull} S.~J.,  {Lavaux} G.,   {Hudson} M.~J.,  2015, \mn@doi
  [\mnras] {10.1093/mnras/stv547}, \href
  {https://ui.adsabs.harvard.edu/abs/2015MNRAS.450..317C} {450, 317}

\bibitem[\protect\citeauthoryear{{Dekel}, {Bertschinger}  \& {Faber}}{{Dekel}
  et~al.}{1990}]{Dekel1990}
{Dekel} A.,  {Bertschinger} E.,   {Faber} S.~M.,  1990, \mn@doi [\apj]
  {10.1086/169418}, \href
  {https://ui.adsabs.harvard.edu/abs/1990ApJ...364..349D} {364, 349}

\bibitem[\protect\citeauthoryear{{Djorgovski} \& {Davis}}{{Djorgovski} \&
  {Davis}}{1987}]{FP2}
{Djorgovski} S.,  {Davis} M.,  1987, \mn@doi [\apj] {10.1086/164948}, \href
  {https://ui.adsabs.harvard.edu/abs/1987ApJ...313...59D} {313, 59}

\bibitem[\protect\citeauthoryear{{Dressler}, {Lynden-Bell}, {Burstein},
  {Davies}, {Faber}, {Terlevich}  \& {Wegner}}{{Dressler} et~al.}{1987}]{FP1}
{Dressler} A.,  {Lynden-Bell} D.,  {Burstein} D.,  {Davies} R.~L.,  {Faber}
  S.~M.,  {Terlevich} R.,   {Wegner} G.,  1987, \mn@doi [\apj]
  {10.1086/164947}, \href
  {https://ui.adsabs.harvard.edu/abs/1987ApJ...313...42D} {313, 42}

\bibitem[\protect\citeauthoryear{{Eddington}}{{Eddington}}{1914}]{eddington14}
{Eddington} A.~S.,  1914, Stellar movements and the structure of the universe

\bibitem[\protect\citeauthoryear{{Folatelli} et~al.}{{Folatelli}
  et~al.}{2010}]{CSP_sn1_folatelli}
{Folatelli} G.,  et~al., 2010, \mn@doi [\aj] {10.1088/0004-6256/139/1/120},
  \href {https://ui.adsabs.harvard.edu/abs/2010AJ....139..120F} {139, 120}

\bibitem[\protect\citeauthoryear{{Foley} et~al.,}{{Foley}
  et~al.}{2018}]{foundation1}
{Foley} R.~J.,  et~al., 2018, \mn@doi [\mnras] {10.1093/mnras/stx3136}, \href
  {https://ui.adsabs.harvard.edu/abs/2018MNRAS.475..193F} {475, 193}

\bibitem[\protect\citeauthoryear{{Ganeshalingam}, {Li}  \&
  {Filippenko}}{{Ganeshalingam} et~al.}{2013}]{LOSS_data}
{Ganeshalingam} M.,  {Li} W.,   {Filippenko} A.~V.,  2013, \mn@doi [\mnras]
  {10.1093/mnras/stt893}, \href
  {https://ui.adsabs.harvard.edu/abs/2013MNRAS.433.2240G} {433, 2240}

\bibitem[\protect\citeauthoryear{{Hicken}, {Wood-Vasey}, {Blondin}, {Challis},
  {Jha}, {Kelly}, {Rest}  \& {Kirshner}}{{Hicken} et~al.}{2009}]{constitution}
{Hicken} M.,  {Wood-Vasey} W.~M.,  {Blondin} S.,  {Challis} P.,  {Jha} S.,
  {Kelly} P.~L.,  {Rest} A.,   {Kirshner} R.~P.,  2009, \mn@doi [\apj]
  {10.1088/0004-637X/700/2/1097}, \href
  {https://ui.adsabs.harvard.edu/abs/2009ApJ...700.1097H} {700, 1097}

\bibitem[\protect\citeauthoryear{{Hong} et~al.,}{{Hong}
  et~al.}{2019}]{2mtf_data}
{Hong} T.,  et~al., 2019, \mn@doi [\mnras] {10.1093/mnras/stz1413}, \href
  {https://ui.adsabs.harvard.edu/abs/2019MNRAS.487.2061H} {487, 2061}

\bibitem[\protect\citeauthoryear{{Howlett} \& {Davis}}{{Howlett} \&
  {Davis}}{2020}]{howlett_davis}
{Howlett} C.,  {Davis} T.~M.,  2020, \mn@doi [\mnras] {10.1093/mnras/staa049},
  \href {https://ui.adsabs.harvard.edu/abs/2020MNRAS.492.3803H} {492, 3803}

\bibitem[\protect\citeauthoryear{{Hudson}}{{Hudson}}{1994a}]{Hudson94a}
{Hudson} M.~J.,  1994a, \mn@doi [\mnras] {10.1093/mnras/266.2.468}, \href
  {https://ui.adsabs.harvard.edu/abs/1994MNRAS.266..468H} {266, 468}

\bibitem[\protect\citeauthoryear{{Hudson}}{{Hudson}}{1994b}]{Hudson94b}
{Hudson} M.~J.,  1994b, \mn@doi [\mnras] {10.1093/mnras/266.2.475}, \href
  {https://ui.adsabs.harvard.edu/abs/1994MNRAS.266..475H} {266, 475}

\bibitem[\protect\citeauthoryear{{Hui} \& {Greene}}{{Hui} \&
  {Greene}}{2006}]{Hui2006}
{Hui} L.,  {Greene} P.~B.,  2006, \mn@doi [\prd] {10.1103/PhysRevD.73.123526},
  \href {https://ui.adsabs.harvard.edu/abs/2006PhRvD..73l3526H} {73, 123526}

\bibitem[\protect\citeauthoryear{{Huterer}, {Shafer}, {Scolnic}  \&
  {Schmidt}}{{Huterer} et~al.}{2017}]{supercal_growth}
{Huterer} D.,  {Shafer} D.~L.,  {Scolnic} D.~M.,   {Schmidt} F.,  2017, \mn@doi
  [\jcap] {10.1088/1475-7516/2017/05/015}, \href
  {https://ui.adsabs.harvard.edu/abs/2017JCAP...05..015H} {2017, 015}

\bibitem[\protect\citeauthoryear{{Jasche} \& {Lavaux}}{{Jasche} \&
  {Lavaux}}{2019}]{borg_pm}
{Jasche} J.,  {Lavaux} G.,  2019, \mn@doi [\aap] {10.1051/0004-6361/201833710},
  \href {https://ui.adsabs.harvard.edu/abs/2019A&A...625A..64J} {625, A64}

\bibitem[\protect\citeauthoryear{{Jasche} \& {Wandelt}}{{Jasche} \&
  {Wandelt}}{2013}]{borg_original}
{Jasche} J.,  {Wandelt} B.~D.,  2013, \mn@doi [\mnras] {10.1093/mnras/stt449},
  \href {https://ui.adsabs.harvard.edu/abs/2013MNRAS.432..894J} {432, 894}

\bibitem[\protect\citeauthoryear{{Jones} et~al.,}{{Jones}
  et~al.}{2019}]{foundation2}
{Jones} D.~O.,  et~al., 2019, \mn@doi [\apj] {10.3847/1538-4357/ab2bec}, \href
  {https://ui.adsabs.harvard.edu/abs/2019ApJ...881...19J} {881, 19}

\bibitem[\protect\citeauthoryear{{Kaiser}, {Efstathiou}, {Saunders}, {Ellis},
  {Frenk}, {Lawrence}  \& {Rowan-Robinson}}{{Kaiser} et~al.}{1991}]{Kaiser91}
{Kaiser} N.,  {Efstathiou} G.,  {Saunders} W.,  {Ellis} R.,  {Frenk} C.,
  {Lawrence} A.,   {Rowan-Robinson} M.,  1991, \mn@doi [\mnras]
  {10.1093/mnras/252.1.1}, \href
  {https://ui.adsabs.harvard.edu/abs/1991MNRAS.252....1K} {252, 1}

\bibitem[\protect\citeauthoryear{{Kodi Ramanah}, {Charnock}  \& {Lavaux}}{{Kodi
  Ramanah} et~al.}{2019}]{halo_painting}
{Kodi Ramanah} D.,  {Charnock} T.,   {Lavaux} G.,  2019, \mn@doi [\prd]
  {10.1103/PhysRevD.100.043515}, \href
  {https://ui.adsabs.harvard.edu/abs/2019PhRvD.100d3515K} {100, 043515}

\bibitem[\protect\citeauthoryear{{Kourkchi} \& {Tully}}{{Kourkchi} \&
  {Tully}}{2017}]{Kourkchi17}
{Kourkchi} E.,  {Tully} R.~B.,  2017, \mn@doi [\apj]
  {10.3847/1538-4357/aa76db}, \href
  {https://ui.adsabs.harvard.edu/abs/2017ApJ...843...16K} {843, 16}

\bibitem[\protect\citeauthoryear{{Krisciunas} et~al.,}{{Krisciunas}
  et~al.}{2017}]{CSP_DR3}
{Krisciunas} K.,  et~al., 2017, \mn@doi [\aj] {10.3847/1538-3881/aa8df0}, \href
  {https://ui.adsabs.harvard.edu/abs/2017AJ....154..211K} {154, 211}

\bibitem[\protect\citeauthoryear{{Lavaux} \& {Hudson}}{{Lavaux} \&
  {Hudson}}{2011}]{2Mpp_paper}
{Lavaux} G.,  {Hudson} M.~J.,  2011, \mn@doi [\mnras]
  {10.1111/j.1365-2966.2011.19233.x}, \href
  {https://ui.adsabs.harvard.edu/abs/2011MNRAS.416.2840L} {416, 2840}

\bibitem[\protect\citeauthoryear{{Lynden-Bell}, {Faber}, {Burstein}, {Davies},
  {Dressler}, {Terlevich}  \& {Wegner}}{{Lynden-Bell}
  et~al.}{1988}]{LyndenBell88}
{Lynden-Bell} D.,  {Faber} S.~M.,  {Burstein} D.,  {Davies} R.~L.,  {Dressler}
  A.,  {Terlevich} R.~J.,   {Wegner} G.,  1988, \mn@doi [\apj]
  {10.1086/166066}, \href
  {https://ui.adsabs.harvard.edu/abs/1988ApJ...326...19L} {326, 19}

\bibitem[\protect\citeauthoryear{{Mackay}}{{Mackay}}{2003}]{mackay_book}
{Mackay} D. J.~C.,  2003, {Information Theory, Inference and Learning
  Algorithms}.
Cambridge University Press

\bibitem[\protect\citeauthoryear{{Magoulas} et~al.,}{{Magoulas}
  et~al.}{2012}]{6df_fp}
{Magoulas} C.,  et~al., 2012, \mn@doi [\mnras]
  {10.1111/j.1365-2966.2012.21421.x}, \href
  {https://ui.adsabs.harvard.edu/abs/2012MNRAS.427..245M} {427, 245}

\bibitem[\protect\citeauthoryear{Malmquist}{Malmquist}{1920}]{Mal20}
Malmquist K.~G.,  1920, Lund Medd. Ser II, 22

\bibitem[\protect\citeauthoryear{{Masters}, {Springob}, {Haynes}  \&
  {Giovanelli}}{{Masters} et~al.}{2006}]{sfi1}
{Masters} K.~L.,  {Springob} C.~M.,  {Haynes} M.~P.,   {Giovanelli} R.,  2006,
  \mn@doi [\apj] {10.1086/508924}, \href
  {https://ui.adsabs.harvard.edu/abs/2006ApJ...653..861M} {653, 861}

\bibitem[\protect\citeauthoryear{{Masters}, {Springob}  \& {Huchra}}{{Masters}
  et~al.}{2008}]{2MTF}
{Masters} K.~L.,  {Springob} C.~M.,   {Huchra} J.~P.,  2008, \mn@doi [\aj]
  {10.1088/0004-6256/135/5/1738}, \href
  {https://ui.adsabs.harvard.edu/abs/2008AJ....135.1738M} {135, 1738}

\bibitem[\protect\citeauthoryear{{Monelli} \& {Trujillo}}{{Monelli} \&
  {Trujillo}}{2019}]{MonTru19}
{Monelli} M.,  {Trujillo} I.,  2019, \mn@doi [\apjl]
  {10.3847/2041-8213/ab2fd2}, \href
  {https://ui.adsabs.harvard.edu/abs/2019ApJ...880L..11M} {880, L11}

\bibitem[\protect\citeauthoryear{{Mukherjee}, {Lavaux}, {Bouchet}, {Jasche},
  {Wandelt}, {Nissanke}, {Leclercq}  \& {Hotokezaka}}{{Mukherjee}
  et~al.}{2019}]{borg_velocity_corrections}
{Mukherjee} S.,  {Lavaux} G.,  {Bouchet} F.~R.,  {Jasche} J.,  {Wandelt} B.~D.,
   {Nissanke} S.~M.,  {Leclercq} F.,   {Hotokezaka} K.,  2019, arXiv e-prints,
  \href {https://ui.adsabs.harvard.edu/abs/2019arXiv190908627M} {p.
  arXiv:1909.08627}

\bibitem[\protect\citeauthoryear{{Neill}, {Hudson}  \& {Conley}}{{Neill}
  et~al.}{2007}]{NeillHudson2007}
{Neill} J.~D.,  {Hudson} M.~J.,   {Conley} A.,  2007, \mn@doi [\apjl]
  {10.1086/518808}, \href
  {https://ui.adsabs.harvard.edu/abs/2007ApJ...661L.123N} {661, L123}

\bibitem[\protect\citeauthoryear{{Nicolaou}, {Lahav}, {Lemos}, {Hartley}  \&
  {Braden}}{{Nicolaou} et~al.}{2019}]{lahav_paper}
{Nicolaou} C.,  {Lahav} O.,  {Lemos} P.,  {Hartley} W.,   {Braden} J.,  2019,
  arXiv e-prints, \href {https://ui.adsabs.harvard.edu/abs/2019arXiv190909609N}
  {p. arXiv:1909.09609}

\bibitem[\protect\citeauthoryear{{Pesce} et~al.,}{{Pesce}
  et~al.}{2020}]{MasersH0}
{Pesce} D.~W.,  et~al., 2020, \mn@doi [\apjl] {10.3847/2041-8213/ab75f0}, \href
  {https://ui.adsabs.harvard.edu/abs/2020ApJ...891L...1P} {891, L1}

\bibitem[\protect\citeauthoryear{{Pike} \& {Hudson}}{{Pike} \&
  {Hudson}}{2005}]{pike_hudson}
{Pike} R.~W.,  {Hudson} M.~J.,  2005, \mn@doi [\apj] {10.1086/497359}, \href
  {https://ui.adsabs.harvard.edu/abs/2005ApJ...635...11P} {635, 11}

\bibitem[\protect\citeauthoryear{{Planck Collaboration} et~al.,}{{Planck
  Collaboration} et~al.}{2018}]{planck_cosmology}
{Planck Collaboration} et~al., 2018, arXiv e-prints, \href
  {https://ui.adsabs.harvard.edu/abs/2018arXiv180706209P} {p. arXiv:1807.06209}

\bibitem[\protect\citeauthoryear{{Riess} et~al.,}{{Riess}
  et~al.}{2011}]{Riess2011}
{Riess} A.~G.,  et~al., 2011, \mn@doi [\apj] {10.1088/0004-637X/730/2/119},
  \href {https://ui.adsabs.harvard.edu/abs/2011ApJ...730..119R} {730, 119}

\bibitem[\protect\citeauthoryear{{Riess}, {Casertano}, {Yuan}, {Macri}  \&
  {Scolnic}}{{Riess} et~al.}{2019}]{shoes_new}
{Riess} A.~G.,  {Casertano} S.,  {Yuan} W.,  {Macri} L.~M.,   {Scolnic} D.,
  2019, \mn@doi [\apj] {10.3847/1538-4357/ab1422}, \href
  {https://ui.adsabs.harvard.edu/abs/2019ApJ...876...85R} {876, 85}

\bibitem[\protect\citeauthoryear{{Springob}, {Masters}, {Haynes}, {Giovanelli}
  \& {Marinoni}}{{Springob} et~al.}{2007}]{sfi2}
{Springob} C.~M.,  {Masters} K.~L.,  {Haynes} M.~P.,  {Giovanelli} R.,
  {Marinoni} C.,  2007, \mn@doi [\apjs] {10.1086/519527}, \href
  {https://ui.adsabs.harvard.edu/abs/2007ApJS..172..599S} {172, 599}

\bibitem[\protect\citeauthoryear{{Springob} et~al.,}{{Springob}
  et~al.}{2014}]{6df_velocity}
{Springob} C.~M.,  et~al., 2014, \mn@doi [\mnras] {10.1093/mnras/stu1743},
  \href {https://ui.adsabs.harvard.edu/abs/2014MNRAS.445.2677S} {445, 2677}

\bibitem[\protect\citeauthoryear{{Springob} et~al.,}{{Springob}
  et~al.}{2016}]{2mtf_cosmography}
{Springob} C.~M.,  et~al., 2016, \mn@doi [\mnras] {10.1093/mnras/stv2648},
  \href {https://ui.adsabs.harvard.edu/abs/2016MNRAS.456.1886S} {456, 1886}

\bibitem[\protect\citeauthoryear{{Strauss} \& {Willick}}{{Strauss} \&
  {Willick}}{1995}]{pec_vel_review}
{Strauss} M.~A.,  {Willick} J.~A.,  1995, \mn@doi [Phys. Rep.]
  {10.1016/0370-1573(95)00013-7}, 261, 271

\bibitem[\protect\citeauthoryear{{Tonry}, {Dressler}, {Blakeslee}, {Ajhar},
  {Fletcher}, {Luppino}, {Metzger}  \& {Moore}}{{Tonry} et~al.}{2001}]{SBFIV}
{Tonry} J.~L.,  {Dressler} A.,  {Blakeslee} J.~P.,  {Ajhar} E.~A.,  {Fletcher}
  A.~B.,  {Luppino} G.~A.,  {Metzger} M.~R.,   {Moore} C.~B.,  2001, \mn@doi
  [\apj] {10.1086/318301}, \href
  {https://ui.adsabs.harvard.edu/abs/2001ApJ...546..681T} {546, 681}

\bibitem[\protect\citeauthoryear{{Trujillo} et~al.,}{{Trujillo}
  et~al.}{2019}]{ngc1052_distance1}
{Trujillo} I.,  et~al., 2019, \mn@doi [\mnras] {10.1093/mnras/stz771}, \href
  {https://ui.adsabs.harvard.edu/abs/2019MNRAS.486.1192T} {486, 1192}

\bibitem[\protect\citeauthoryear{{Tully} \& {Fisher}}{{Tully} \&
  {Fisher}}{1977}]{TF_relations}
{Tully} R.~B.,  {Fisher} J.~R.,  1977, \aap, \href
  {https://ui.adsabs.harvard.edu/abs/1977A&A....54..661T} {500, 105}

\bibitem[\protect\citeauthoryear{{Verde}, {Treu}  \& {Riess}}{{Verde}
  et~al.}{2019}]{H0_tension_review}
{Verde} L.,  {Treu} T.,   {Riess} A.~G.,  2019, \mn@doi [Nature Astronomy]
  {10.1038/s41550-019-0902-0}, \href
  {https://ui.adsabs.harvard.edu/abs/2019NatAs...3..891V} {3, 891}

\bibitem[\protect\citeauthoryear{{Wong} et~al.,}{{Wong} et~al.}{2019}]{H0licow}
{Wong} K.~C.,  et~al., 2019, arXiv e-prints, \href
  {https://ui.adsabs.harvard.edu/abs/2019arXiv190704869W} {p. arXiv:1907.04869}

\bibitem[\protect\citeauthoryear{{van Dokkum} et~al.,}{{van Dokkum}
  et~al.}{2018}]{ngc1052_discovery}
{van Dokkum} P.,  et~al., 2018, \mn@doi [\nat] {10.1038/nature25767}, \href
  {https://ui.adsabs.harvard.edu/abs/2018Natur.555..629V} {555, 629}

\makeatother
\end{thebibliography}
\appendix
\section{Tests with simulations}\label{sec:KS_simulation}

\begin{figure*}
    \centering
    \includegraphics[width=\linewidth]{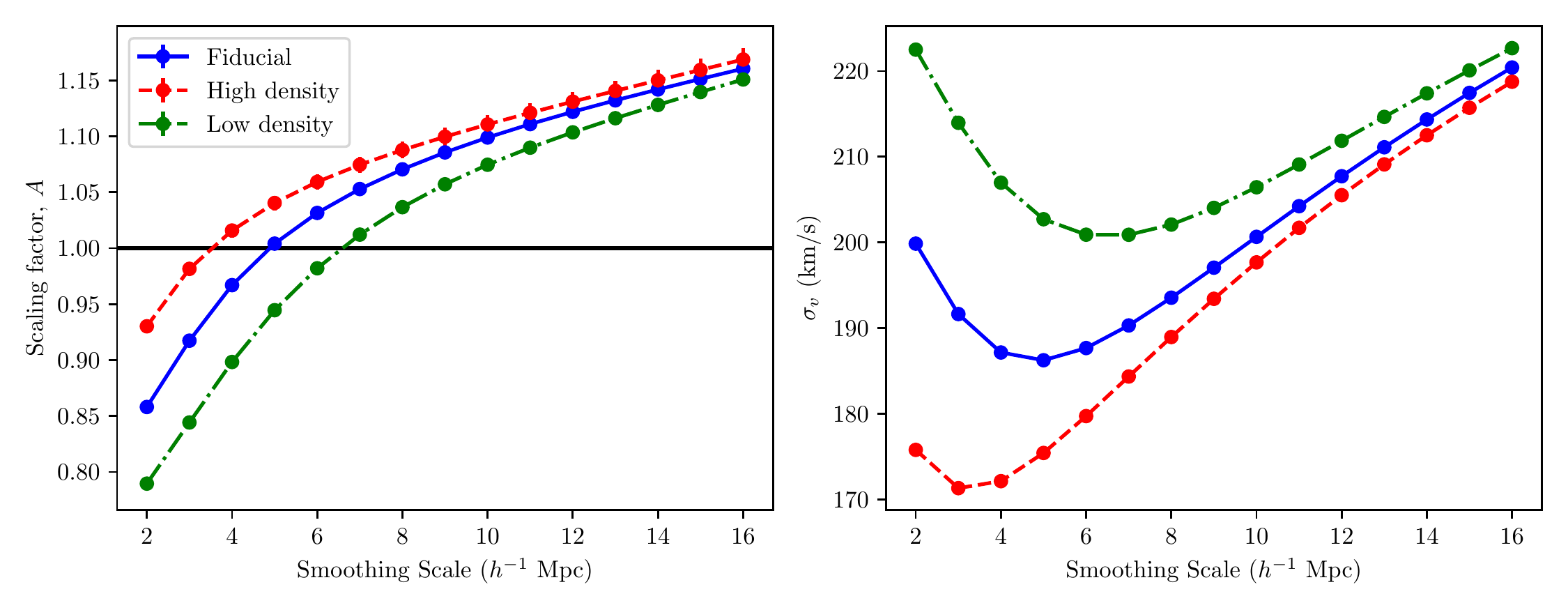}
    \caption{The dependence of the \textit{(left)} scaling factor, and \textit{(right)} the velocity error, on the smoothing scale in our simulations. The blue, green and the red curves show the results for the fiducial, high and low tracer density samples. For the scaling factor, we also plot the results from the linear theory. As can be seen from the figure, the velocity error is lower for a high tracer density. Also note that the scale for unbiased velocity prediction depends on the tracer density of the sample. For the sample with high tracer density, the smoothing scale of unbiased velocity predictions is lower.}
    \label{fig:KS_sim}
\end{figure*}

In Section \ref{sec:scaling}, we used a value of $R_{\text{unbiased}}$, which denotes the smoothing scale at which the predictions from kernel smoothing of peculiar velocities give an unbiased result. In this section, we determine the value of this scale using an N-body simulation. We used a N-body simulation from the VELMASS simulation suite\footnote{See \citet{halo_painting} for more details on the simulation}. The VELMASS simulation we use was performed in a cubic box of size $2~h^{-1}$ Gpc with a total of $2048^3$ particles with mass $9.387\times 10^{10}~h^{-1}M_{\odot}$. The cosmological parameters for this simulation are as follows: $\Omega_m= 0.315, \Omega_b = 0.049, H_0 = 68$ km s$^{-1}$ Mpc$^{-1}$, $\sigma_8 = 0.81$, $n_s = 0.97$ and $Y_{\text{He}} = 0.248$. We identified the halos in the simulation with the ROCKSTAR halo finding software \citep{rockstar}, using only the halos of mass larger than $2\times10^{12}~h^{-1}M_{\odot}$. We used the halos within a ($250~h^{-1}$ Mpc)$^3$ sub-box of the full simulation. We do not consider the sub-halos. There were a total of $32784$ halos in this region. We then split these halos into two sets - a `tracer' sample which is used to obtain the kernel smoothed velocity and a `test' sample, for which the predicted velocity is compared to the true velocity. 

We use three different number densities for the tracer population to validate the prediction. In the fiducial configuration, we used $12000$ randomly selected tracer velocities to predict the velocity. This roughly corresponds to the tracer density of the 2MTF survey. In addition to this fiducial tracer density, we also used a high (low) tracer density configuration with $24000$ ($6000$) randomly selected tracers. For each of these tracer densities, we compared the predicted velocity of the test halos to its true velocity, thus fitting for the scaling factor, $A$ and the velocity error, $\sigma_v$. The scaling factor scales the predicted velocity such that, 
\begin{equation}
    \mvec{V}_{\text{true}} = A \mvec{V}_{\text{pred}}.
\end{equation}
Also note that, for the simulations, we are smoothing the full 3-dimensional velocity and not just the radial velocity. The value of the $A$ and $\sigma_v$ is fitted for different smoothing scales. The results of these fits are shown in Figure \ref{fig:KS_sim}. As expected, the velocity error, $\sigma_v$ is larger for the sample with low tracer density. For each of the three tracer densities, the unbiased smoothing length is around $3$-$5~h^{-1}$ Mpc. There is, nevertheless, a dependence of this scale on the tracer density - for a higher tracer density, the unbiased smoothing scale is lower. A similar trend is also obtained for the scale where the velocity error is minimum (See the right panel of Figure \ref{fig:KS_sim}). 
\section{Results for the posterior ratio}\label{sec:posterior_ratio}

In section \ref{ssec:fwd_lkl}, we used Bayesian model comparison to compare the reconstructed peculiar velocity field from 2M++ with the adaptive kernel-smoothed velocity fields. In doing so, we compared the posterior ratios using the forward likelihood presented in the same section. We present the values of the posterior ratios in Tables \ref{tbl:fwd_lkl} and \ref{tbl:fwd_lkl_16}, where we compared the adaptive kernel-smoothed velocity fields smoothed with a fiducial smoothing length of $8~h^{-1}$ Mpc and $16~h^{-1}$ Mpc respectively. The reported quantity is the ratio of likelihood obtained for the 2M++ field and the adaptive kernel smoothed field. We make various cuts in redshifts as indicated in the tables. We compare the likelihoods both with (denoted as a superscript) and without the scaling factor introduced in section \ref{sec:scaling}. In the following tables, a positive value for the posterior ratio implies that the 2M++ velocity field is preferred.

\begin{table*}
  \centering
  \caption{Ratio of $\log(\mP)$ calculated using the 2M++ reconstructed velocity field and the adaptive kernel smoothing technique for 6dF and SuperTF. For this table, we used a fixed value of $\sigma_v = 150$ km s$^{-1}$ and the fiducial smoothing length is $8~h^{-1}$ Mpc. For each test set, we make cuts in the redshift as indicated. The prefix `scaled' denotes that the kernel-smoothed velocity is scaled up by a scaling factor. The scaling factor for $8~h^{-1}$ Mpc smoothing is $1.07$.}
  \begin{tabular}{l l c c c c c}
  \hline

     Test set & Redshift selection &  $\ln\bigg(\frac{\mP_{\text{2M++}}}{\mP_{\text{6dF}}}\bigg)$ & $\ln\bigg(\frac{\mP_{\text{2M++}}}{\mP^{\text{scaled}}_{\text{6dF}}}\bigg)$ & $\ln\bigg(\frac{\mP_{\text{2M++}}}{\mP_{\text{SuperTF}}}\bigg)$ &     $\ln\bigg(\frac{\mP_{\text{2M++}}}{\mP^{\text{scaled}}_{\text{SuperTF}}}\bigg)$ &$N_{\text{tracers}}$\\
    \hline
    A2-South &  $cz < 3000$ km/s & $2.21$ & $2.70$ & $2.41$ & $2.88$ & $16$ \\
    & $cz < 4500$ km/s& $2.33$ & $3.86$ & $5.07$ & $5.89$ & $32$ \\
    & $cz < 6000$ km/s& $5.21$ & $6.05$ & $8.85$ & $10.21$ & $53$ \\
    & $cz < 9000$ km/s& $9.19$ & $10.58$ & $15.56$ & $17.80$ & $79$ \\
    \hline
    A2-low-$z$ &  $cz < 3000$ km/s & --- & --- & $11.38$ & $11.95$ & $49$ \\
    & $cz < 4500$ km/s& --- & --- & $20.76$ & 21.89 & $92$ \\
    & $cz < 6000$ km/s& --- & --- & 49.82 & 52.86 & 168 \\
    & $cz < 9000$ km/s& --- & --- & 85.92 & 92.90 & 310 \\
    \hline
    2MTF &  $cz < 3000$ km/s & 24.6 & 24.4 & --- & --- & 108 \\
    & $cz < 4500$ km/s& 39.00 & 36.28 & --- & --- & 247 \\
    & $cz < 6000$ km/s& 55.65 & 53.06 & --- & --- & 379 \\
    & $cz < 9000$ km/s& 69.49 & 69.01 & --- & --- & 483 \\
    \hline
    SFI++ Groups &  $cz < 3000$ km/s & 12.08 & 11.88 & --- & --- & 61 \\
    & $cz < 4500$ km/s& 9.35 & 9.29 & --- & --- & 100 \\
    & $cz < 6000$ km/s& 18.89 & 17.87 & --- & --- & 165 \\
    & $cz < 9000$ km/s& 18.78 & 17.78 & --- & --- & 170 \\
    \hline
    SFI++ Field &  $cz < 3000$ km/s & 9.94 & 9.01 & --- & --- & 63 \\
    & $cz < 4500$ km/s& 5.72 & 5.24 & --- & --- & 153 \\
    & $cz < 6000$ km/s& 13.52 & 11.70 & --- & --- & 388 \\
    & $cz < 9000$ km/s& 53.66 & 44.29 & --- & --- & 736 \\
    \hline
  \end{tabular}
  \label{tbl:fwd_lkl}
\end{table*}

\begin{table*}
  \centering
  \caption{Same as table \ref{tbl:fwd_lkl}, but with a fiducial smoothing length of $16~h^{-1}$ Mpc. The scaling factor for $16~h^{-1}$ Mpc smoothing is 1.16.}
  \begin{tabular}{l l c c c c c}
  \hline
     Test set & Redshift selection &  $\ln\bigg(\frac{\mP_{\text{2M++}}}{\mP_{\text{6dF}}}\bigg)$ & $\ln\bigg(\frac{\mP_{\text{2M++}}}{\mP^{\text{scaled}}_{\text{6dF}}}\bigg)$ & $\ln\bigg(\frac{\mP_{\text{2M++}}}{\mP_{\text{SuperTF}}}\bigg)$ &     $\ln\bigg(\frac{\mP_{\text{2M++}}}{\mP^{\text{scaled}}_{\text{SuperTF}}}\bigg)$ &$N_{\text{tracers}}$\\
    \hline
    A2-South &  $cz < 3000$ km/s & $2.39$ & $2.61$ & $3.01$ & $3.11$ & $23$ \\
    & $cz < 4500$ km/s& $4.37$ & $4.48$ & $6.19$ & $6.47$ & $42$ \\
    & $cz < 6000$ km/s& $6.18$ & $6.33$ & $9.70$ & $10.08$ & $66$ \\
    & $cz < 9000$ km/s& $8.01$ & $8.05$ & $12.83$ & $12.83$ & $94$ \\
    \hline
    A2-low-$z$ &  $cz < 3000$ km/s & --- & --- & $9.96$ & $9.56$ & $49$ \\
    & $cz < 4500$ km/s& --- & --- & $17.50$ & $16.80$ & $92$ \\
    & $cz < 6000$ km/s& --- & --- & $42.78$ & $42.42$ & 168 \\
    & $cz < 9000$ km/s& --- & --- & $69.66$ & $68.81$ & 310 \\
    \hline
    2MTF &  $cz < 3000$ km/s & 20.81 & 18.51 & --- & --- & 118 \\
    & $cz < 4500$ km/s& 45.63 & 39.00 & --- & --- & 282 \\
    & $cz < 6000$ km/s& 59.87 & 51.25 & --- & --- & 443 \\
    & $cz < 9000$ km/s& 68.19 & 59.85 & --- & --- & 563 \\
    \hline
    SFI++ Groups &  $cz < 3000$ km/s & 8.36 & 7.18 & --- & --- & 70 \\
    & $cz < 4500$ km/s& 7.22 & 5.17 & --- & --- & 119 \\
    & $cz < 6000$ km/s& 10.19 & 7.56 & --- & --- & 198 \\
    & $cz < 9000$ km/s& 9.99 & 7.33 & --- & --- & 203 \\
    \hline
    SFI++ Field &  $cz < 3000$ km/s & 8.93 & 9.32 & --- & --- & 75 \\
    & $cz < 4500$ km/s& 5.16 & 4.58 & --- & --- & 180 \\
    & $cz < 6000$ km/s& 9.40 & 7.14 & --- & --- & 450 \\
    & $cz < 9000$ km/s& 10.87 & 10.86 & --- & --- & 863 \\
    \hline
  \end{tabular}
  \label{tbl:fwd_lkl_16}
\end{table*}




\bsp	
\label{lastpage}
\end{document}